\documentclass[aps,twocolumn,floats,prd,nofootinbib,10pt,longbibliography]{revtex4-1}
\pdfoutput=1
\usepackage{subfigure}
\usepackage{amsfonts}
\usepackage{amsmath}
\usepackage[utf8]{inputenc}
\usepackage[british]{babel}
\usepackage[Symbolsmallscale]{upgreek}
\usepackage[protrusion=true,tracking=true,kerning=true,spacing=true,final,babel=true]{microtype}
\usepackage{physics}
\usepackage{xcolor}
\usepackage{graphicx}
\usepackage[final]{hyperref}
\usepackage{nth}

\newcommand{ \Planck}{{\it Planck}}
\newcommand{\LCDM}{$\Lambda$CDM}
\newcommand{\SHOES}{S$H_0$ES}
\def\be{\begin{equation}}
\newcommand{\lsim}{\mathrel{\hbox{\rlap{\lower.55ex\hbox{$\sim$}} \kern-.3em \raise.4ex \hbox{$<$}}}}
\newcommand{\gsim}{\mathrel{\hbox{\rlap{\lower.55ex\hbox{$\sim$}} \kern-.3em \raise.4ex \hbox{$>$}}}}

\def\ee{\end{equation}}

\begin{document}

\title{Current small-scale CMB constraints to axion-like early dark energy} 

\author{Tristan L.~Smith}
\affiliation{Department of Physics and Astronomy, Swarthmore College, \\ 500 College Ave., Swarthmore, PA 19081, USA}
\affiliation{Center for Cosmology and Particle Physics, Department of Physics, New York University, New York, NY 10003, USA}

\author{Vivian Poulin}
\affiliation{Laboratoire Univers \& Particules de Montpellier (LUPM), CNRS \& Universit\'{e} de Montpellier (UMR-5299), Place Eug\`{e}ne Bataillon, F-34095 Montpellier Cedex 05, France}

\date{\today}

\begin{abstract}
 The SPT-3G 2018 TT/TE/EE cosmic microwave background (CMB) data set (temperature and polarization) is used to place constraints on an axion-like model of early dark energy (EDE). These data do not favor axion-like EDE and place an upper limit on the maximum fraction of the total energy density $f_{\rm EDE}< 0.172$ (at the 95\% confidence level, CL). This is in contrast with ACT DR4 which gives $f_{\rm EDE}=0.150^{+0.050}_{-0.078}$. When combining CMB measurements with measurements of the baryon acoustic oscillations and luminosity distance to Type Ia supernovae, we show that the tension with the S$H_0$ES measurement of the Hubble parameter goes up from 2.6$\sigma$ with \textit{Planck} to 2.9$\sigma$ with \textit{Planck}+SPT-3G 2018. The additional inclusion of ACT DR4 data leads to a reduction of the tension to $1.6\sigma$, but the discrepancy between ACT DR4 and {\it Planck}+SPT-3G 2018 casts some doubt on the statistical consistency of this joint analysis. The importance of improved measurements of the CMB at both intermediate and small scales (in particular the shape of the damping tail) as well as the interplay between temperature and polarization measurements in constraining EDE are discussed. Upcoming ground-based measurements of the CMB will play a crucial role in determining whether EDE  remains a viable model to address the Hubble tension. 
\end{abstract}

\maketitle

\section{Introduction}

Since the turn of the millennium, we have been living in the age of `precision cosmology' \cite{Turner:2022gvw}. Measurements of the cosmic microwave background (CMB), the clustering of large scale structure (LSS)-- and in particular the baryon acoustic oscillations (BAO), type Ia supernovae (SNeIa), and the primordial abundance of light elements produced during big bang nucleosynthesis (BBN), have largely confirmed the core cosmological model. This model consists of baryons, photons, neutrinos, cold dark matter, and a cosmological constant ($\Lambda$), i.e., `$\Lambda$CDM'. By performing fits to a suite of high precision data sets, we are able to obtain percent-level precision in estimates of the values of the six free cosmological parameters of the models (see, e.g., Ref.~\cite{2020RvMP...92c0501P}).  

As our measurements have become increasingly sensitive, a few hints of potential cracks in $\Lambda$CDM have recently appeared. The most significant of these is a mismatch between `direct' (i.e.~kinematical) measurements of the current expansion rate-- known as the Hubble constant, $H_0$-- and the `indirect' (i.e.~dynamical) measurements of $H_0$ inferred through observations that depend on a detailed model of the cosmological dynamics. For a flat $\Lambda$CDM cosmology, using Cepheid variable calibrated SNeIa absolute luminosities (i.e., \SHOES\ \cite{Brout:2022vxf}) and the value of $H_0$ inferred from \textit{Planck} \cite{Planck:2018vyg} gives a $\sim 10\%$ discrepancy with a $\sim 5 \sigma$ statistical significance. Other indirect probes, such as measurements of the BAO, are consistent with the value of $H_0$ inferred from CMB data. There is a larger spread of values from various direct probes, but all of them are larger than those from indirect probes (see, e.g., Ref.~\cite{Riess:2023egm}). Intense  experimental efforts are making it increasingly unlikely that a single source of systematic error could be responsible for these discrepancies (see e.g. Ref.~\cite{Riess:2023bfx} for a recent discussion). This clearly motivates the need to look for a possible explanation of this tension via some physics beyond $\Lambda$CDM, with the wealth of high-precision cosmological data at our disposal.

Several extensions of \LCDM\ which address the Hubble tension have been proposed (for reviews see Refs.~\cite{Schoneberg:2021qvd,DiValentino:2021izs}). One model which has stood out is an axion-like early dark energy (EDE) \cite{Karwal:2016vyq,Poulin:2018cxd,Smith:2019ihp}. This model augments \LCDM\ with a cosmological scalar field which is initially held fixed in its potential by Hubble friction, becomes dynamical around matter-radiation equality, and then dilutes faster than matter. The presence of this field briefly increases the Hubble parameter leading to a decrease in the sound horizon which, in turn, increases the value of $H_0$ inferred from CMB and BAO data. For a thorough review of the original proposal and subsequent improvements and analyses, we refer to Refs.~\cite{Kamionkowski:2022pkx,Poulin:2023lkg}.

\begin{figure}
    \centering
    \includegraphics[width=0.8\columnwidth]{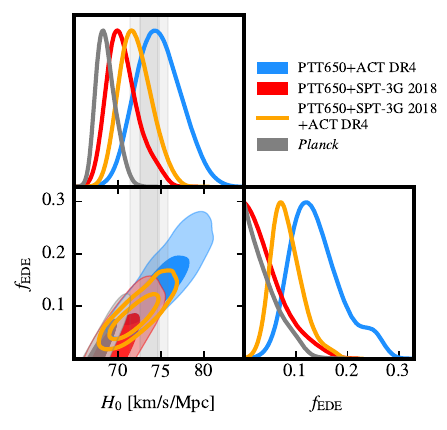}
    \caption{A triangle plot summarizing our main results. The combination of the \textit{Planck} temperature power spectrum restricted to multipoles $\ell \leq 650$ (`PTT650', which is statistically equivalent to WMAP \cite{Huang:2018xle}) and SPT-3G 2018 limits EDE to nearly the same extent as the full \textit{Planck} data set. This is in contrast with ACT DR4 which shows a strong preference for EDE. The combination of PTT650+SPT-3G 2018+ACT DR4 is shown in orange. The gray bands correspond to the S$H_0$ES + Pantheon+  determination of the Hubble constant \cite{Brout:2022vxf}.}
    \label{fig:small}
\end{figure}

Past investigations of EDE with CMB data have led to a mixed picture: on the one hand, \textit{Planck} CMB measurements place an upper limit on the EDE energy density with a correspondingly small change to the posterior distribution for the Hubble constant ($H_0 = 67.34_{-0.65}^{+0.59}\ {\rm km/s/Mpc}\rightarrow H_0 = 68.51_{-1.4}^{+0.76}\ {\rm km/s/Mpc}$).  On the other hand, CMB measurements from ACT DR4 (temperature and polarization), alone or in combination with WMAP, \textit{Planck} polarization and SPT-3G 2018 polarization data lead to $H_0 = 74.2^{+1.9}_{-2.1}$ km/s/Mpc with a $\gtrsim 3 \sigma$ preference for EDE~\cite{Smith:2022hwi}.  The inclusion of the full \textit{Planck} temperature power spectrum moves the inferred value of $H_0$ nearly back to its \LCDM\ value, and the contribution of EDE is compatible with zero at $1\sigma$. However, previous work has shown that part of the apparent constraining power from \textit{Planck} is due to prior volume effects \cite{Smith:2020rxx,Herold:2021ksg,Herold:2022iib}.  
The difference between analyses of \textit{Planck} and ACT DR4 motivates further investigation with an independent CMB data set, such as SPT-3G 2018.

Since these previous analyses were published, the SPT-3G 2018 temperature likelihood was made public \cite{SPT-3G:2022hvq}. Here we explore how the SPT-3G 2018 temperature power spectrum constrains EDE.\footnote{A recent study \cite{FrancoAbellan:2023gec} performed an analysis of a model of Early Modified Gravity (EMG) with some similarities to the EDE model in light of the same datasets. Ref.~\cite{FrancoAbellan:2023gec} reports a preference for EMG at $\sim2\sigma$ in a combined analysis of Planck+SPT-3G 2018+ACT DR4 driven (mostly) by ACT DR4, but a residual $3\sigma$ tension with S$H_0$ES.} 
Our main result is shown in Fig.~\ref{fig:small}, where we display the posterior distributions for the Hubble constant, $H_0$, and the maximum fraction of the total energy density in EDE, $f_{\rm EDE}$. There we can see that both \textit{Planck} and PTT650+SPT-3G 2018\footnote{`PTT650' refers to the \textit{Planck} temperature power spectrum restricted to $\ell \leq 650$. This subset of the full \textit{Planck} data set is statistically equivalent to WMAP \cite{Huang:2018xle}.} show no preference for EDE, whereas PTT650+ACT DR4 shows a significant preference \cite{Hill:2021yec,Poulin:2021bjr,Smith:2022hwi}. Taken at face value, it supports the idea that the hint of EDE in ACT DR4 may be a statistical fluctuation, or a systematic error. The combination of ACT DR4 and SPT-3G 2018 data reduces the preference for EDE over $\Lambda$CDM, when compared to ACT DR4 alone.

The rest of the paper is organized as follows: In Sec.~\ref{sec:method} we describe our analysis setup and the various data sets we have used. In Sec.~\ref{sec:results} we present constraints from \textit{Planck}, ACT DR4, and SPT-3G 2018 on both $\Lambda$CDM and EDE, and highlight the role of the small angular scale measurements of the CMB power spectra in breaking parameter degeneracies. We also explore constraints on EDE from TT and TE/EE separately, finding that when taken individually, they lead to no significant constraints on EDE, but exhibit a mild disagreement at the $\sim2.5\sigma$ level, at the origin of the constraints on EDE from SPT. In Sec.~\ref{sec:ext}, we include non-CMB data sets, and obtain the most up-to-date constraints to EDE from a combination of cosmological data and quantify the ability for EDE to resolve the Hubble tension when using the different CMB data sets. We give our conclusions in Sec.~\ref{sec:concl}. App.~\ref{app:old} provides a comparison between new and old SPT-3G 2018 results. All relevant $\chi^2$ statistics and additional triangles plots are provided in App.~\ref{app:tables}.

Note that for the rest of the paper we use the `reduced' Hubble parameter, $h\equiv H_0/(100\ {\rm km/s/Mpc})$.

\section{Analysis method and data sets}\label{sec:method}

To evaluate the cosmological constraints we perform a series of Markov-chain Monte Carlo (MCMC) runs using either {\sf MontePython-v3}\footnote{\url{https://github.com/brinckmann/montepython_public}} \citep{Audren:2012wb,Brinckmann:2018cvx} or {\sf CosmoMC}\footnote{\url{https://github.com/cmbant/CosmoMC}}, interfaced with versions of either {\sf CLASS}\footnote{\url{https://lesgourg.github.io/class_public/class.html}} \cite{Lesgourgues:2011re,Blas:2011rf} or {\sf CAMB}, respectively, which have been modified to solve for the dynamics of an oscillating cosmological scalar field. {\sf CosmoMC} was used only when analyzing the SPT-3G 2018 temperature and polarization separately.
We have confirmed that the EDE CMB power spectra computed in {\sf CAMB} and {\sf CLASS} agree to better than a fractional difference of $0.001$. We make use of a Metropolis-Hasting algorithm and for analyses that include {\em Planck} large-scale measurements of the E-mode polarization  we use uninformative flat priors on $\{\omega_b,\omega_{\rm cdm},h,\ln(10^{10}A_s),n_s,\tau_{\rm reio}\}$; for analyses that do not include the \textit{Planck} large-scale CMB E-mode power spectrum we use a Gaussian prior on $\tau_{\rm reio} = 0.0540\pm 0.0074$ \cite{SPT-3G:2022hvq}.\footnote{Here $\omega_b \equiv \Omega_b h^2$ and $\omega_{\rm cdm} \equiv \Omega_m h^2$ are the physical baryon and cold dark matter energy densities, respectively, $A_s$ is the amplitude of the scalar perturbations, $n_s$ is the scalar spectral index, and $\tau_{\rm reio}$ is the optical depth to reionization.}

We adopt the {\em Planck} collaboration convention in modeling free-streaming neutrinos as two massless species and one massive with $m_\nu=0.06$ eV \cite{Planck:2018vyg} and use the standard pivot scale, $k_p \equiv 0.05\ {\rm Mpc}^{-1}$. We use {\sf Halofit} to estimate the non-linear matter clustering \cite{Smith:2002dz}. We consider chains to be converged using the Gelman-Rubin \citep{Gelman:1992zz} criterion $|R -1|\lesssim0.05$.\footnote{This condition is chosen because of the non-Gaussian (and sometimes multi-modal) shape of the posteriors of the parameters. For all \LCDM{} runs we have $|R -1|<0.01$.} To analyze the chains and produce our figures we use {\sf GetDist} \cite{Lewis:2019xzd}, and we obtain the minimal $\chi^2$ values using the same method as employed in Ref.~\cite{Schoneberg:2021qvd}.

We make use of the following likelihoods:
\begin{itemize}
    \item \textbf{Planck:} The {\sf Plik} low-$\ell$ CMB temperature and polarization auto-correlations (TT, EE), and the high-$\ell$ TT/TE/EE data~\cite{Planck:2019nip}. In some analyses we combine ground-based CMB measurements with a subset of the \textit{Planck} TT power spectrum with $\ell \leq 650$, which we denote by `PTT650'.  This subset of the \textit{Planck} data has been shown to be in statistical agreement with the Wilkinson Microwave Anisotropy Probe (WMAP) \cite{Huang:2018xle}. We take this agreement between two independent instruments/pipelines as evidence that this subset of the data has negligible systematic errors. When assessing the tension between different data sets we include the gravitational lensing potential reconstruction from {\it Planck}~2018~\cite{Planck:2018lbu}.  
    \item \textbf{SPT-3G 2018:} The most recent SPT-3G 2018 TT/TE/EE likelihood \cite{SPT-3G:2022hvq} which includes temperature and polarization power spectra.\footnote{\url{https://pole.uchicago.edu/public/data/balkenhol22/}} When computing the temperature/polarization-only SPT-3G 2018 constraints we use the original likelihood which is incorporated into {\sf CosmoMC} along with a version of {\sf CAMB} which solves for the dynamics of EDE. When using the full SPT-3G 2018 data set we use the likelihood which has been adapted into the {\sf clik} format paired with {\sf MontePython} format\footnote{\url{https://github.com/SouthPoleTelescope/spt3g_y1_dist}}. In order to compare with previous results we also use the previous SPT-3G 2018 TE/EE release \cite{SPT-3G:2021eoc}  which has been adapted into the {\sf clik} format paired with {\sf MontePython} format\footnote{\url{https://pole.uchicago.edu/public/data/dutcher21}}.
    \item \textbf{ACT DR4:} The ACT DR4  \cite{ACT:2020frw} TT/TE/EE likelihood \footnote{\url{https://github.com/ACTCollaboration/pyactlike}}. In analyses that include the full \textit{Planck} TT power spectrum, we removed any overlap with ACT DR4 TT up until $\ell = 1800$ to avoid introducing correlations between the two data sets \cite{ACT:2020gnv}.
    \item \textbf{BAO:} BAO data from SDSS DR7 at $z = 0.15$ \cite{Ross:2014qpa} and BOSS DR12 at ${z = 0.38, 0.51, 0.61}$ \cite{BOSS:2016wmc}.
    \item \textbf{Pantheon+:} The Pantheon+ catalog of uncalibrated luminosity distance of type Ia supernovae (SNeIa) in the range ${0.01<z<2.26}$ \cite{Brout:2022vxf}. 
    \item $\boldsymbol{M_b}$: A Gaussian prior from the late-time measurement of the absolute calibration of the SNeIa from S$H_0$ES, $M_b = -19.253 \pm 0.027$ \cite{Riess:2021jrx}, corresponding to $H_0 = (73.04\pm1.04)$ km/s/Mpc in $\Lambda$CDM.
\end{itemize}

The `axion-like' EDE model consists of a minimally coupled cosmological scalar field, $\phi$, with a canonical kinetic term and a potential of the form \cite{Smith:2019ihp}
\begin{equation}
    V(\phi) = m^2 f^2 \left(1-\cos \phi/f\right)^3.
\end{equation}
When constraining the EDE cosmology we vary three additional parameters: the logarithm of the redshift at which the EDE component contributes its maximum fraction of the total energy density, $\log_{10}z_c \in [3,4]$, the value of this maximum fraction, $f_{\rm EDE} \equiv \rho_{\rm EDE}(z_c)/\rho_{\rm tot}(z_c) \in [0,0.5]$, and the initial value of the EDE field value, $\phi_i/f  \equiv \theta_i\in [0,3.1]$. We use a shooting algorithm to take the values of $\log_{10}z_c$ and $f_{\rm EDE}$ to find the associated values of $m$ and $f$. The accuracy settings are chosen to ensure that we resolve the oscillations in the field value in both the background and perturbations. 

\section{Constraints from \textit{Planck}, ACT DR4, and SPT-3G 2018}
\label{sec:results}

Measurements of the CMB power spectra give us exquisite information about the acoustic oscillations in the tightly coupled photon-baryon fluid before the photons decoupled \cite{Hu:2000ti}: the angular `wavelength' tells us the angular size of the acoustic horizon at photon decoupling ($\theta_s$), the relative heights of the peaks tell us the relative density of baryons ($\omega_b$) and cold dark matter ($\omega_{cdm}$), the broadband shape tells us the overall amplitude ($A_s$) and slope ($n_s$) of the primordial curvature perturbations, the angular size of the horizon at matter/radiation equality ($\theta_{\rm eq}$), and the angular size of the scale at which photon diffusion causes perturbations to damp away ($\theta_D$, i.e. the `Silk' damping tail) \cite{Silk:1967kq}. 

Let us recall that the key angular scales at play, namely the angular size of the sound horizon $\theta_s$ and the diffusion scale at recombination $\theta_D$, are computed according to the \textit{Planck} collaboration's conventions \cite{Planck:2013pxb}:
\begin{eqnarray}
\theta_s &\equiv& \frac{r_s(z_*)}{D_A(z_*)},\\
r_s(z_*) &=& \int_{z_*}^{\infty} \frac{dz'}{H(z') \sqrt{3(1+R)}},\label{eq:rs}\\
D_A(z_*) &=& \frac{1}{1+z_*}\int_0^{z_*} \frac{dz'}{H(z')}, \\
\theta_D(z_*) &\equiv&\frac{\pi}{k_D(z_*) D_A(z_*)}, \label{eq:thetaD}\\
k^{-2}_D &\equiv& -\frac{1}{6}  \int_{z_*}^{\infty}\frac{dz'}{\dot \tau H(z')}\frac{R^2+16(1+R)/15}{(1+R)^2} 
\end{eqnarray}
where $z_*$ is the redshift at recombination, $R \equiv 3 \rho_b/(4\rho_\gamma)$, and the rate of change of the photon's optical depth can be written $\dot \tau = n_e \sigma_T a$, where $n_e$ is the free electron fraction and $\sigma_T$ is the Thomson scattering cross section. 
From these equations it is clear that in the EDE cosmology the presence of additional energy density pre-recombination, which boosts $H(z)$, directly impacts the sound horizon and damping scale. In addition, the non-zero equation of state and sound speed of the EDE component prevents it from clustering, in turn suppressing the growth of perturbations in the CDM \cite{Poulin:2023lkg}.

The CMB has been observed from both satellites and from ground-based observatories. The most precise measurements come from the \textit{Planck} satellite, which extend to angular scales $\sim 0.07^\circ$ (multipoles around $2\leq \ell \lesssim 2500$). Ground-based measurements from the ACT and SPT collaborations have higher angular resolution, measuring angular scales up to $\sim 0.04^\circ$ ($300\leq \ell \lesssim 4000$). For the angular scales which overlap between \textit{Planck} and these ground-based observatories we gain independent measurements with different systematic uncertainties, for those smaller scales only accessible to the ground-based observatories we gain information about the damping tail as well as a larger lever arm with which to estimate the slope of the primordial curvature perturbations. 

In the following discussion we will take the independent cosmological parameters to be $\omega_{cdm}$, $\omega_b$, $A_s$, $n_s$, $\theta_s$, and $\tau_{\rm reio}$. Since $\theta_s$ is so well measured from the data when we compute parameter degeneracies we fix it to its $\Lambda$CDM \textit{Planck} best fit value $100 \theta_s=1.041085$ \cite{Planck:2018vyg}.  

\subsection{Constraints on $\Lambda$CDM}

Within $\Lambda$CDM there is an important complementarity between intermediate scale measurements of the CMB which do not include information about the damping tail (i.e., $\ell \lesssim 1000$) and measurements which extend to smaller scales (e.g., Ref.~\cite{Addison:2021amj}).  

Requiring that the shape of the damping tail remains relatively unchanged, one obtains the correlation
\begin{eqnarray}
    \frac{\delta \theta_D}{\theta_D} &\simeq& 0.2 \frac{\delta n_s}{n_s}\,.\label{eq:small1}
\end{eqnarray}
This can be simply understood by noting that an increase in $\theta_D$ causes the damping to start on larger scales leading to a decrease in the small-scale amplitude; similarly, for $\ell\gtrsim500$ (i.e., $k>k_p = 0.05\ {\rm Mpc}^{-1}$) an increase in $n_s$ leads to an increase in the small-scale amplitude. This implies that $\theta_D$ and $n_s$ will be positively correlated (see also Ref.~\cite{Addison:2021amj}).
In addition we can use Eq.~(\ref{eq:thetaD}) to relate $\theta_D$ to \LCDM{} parameters:
\begin{eqnarray}
     \frac{\delta \theta_D}{\theta_D} &\simeq& -0.2 \frac{\delta \omega_b}{\omega_b}-0.015 \frac{\delta \omega_{cdm}}{\omega_{cdm}}\,.\label{eq:small2}
\end{eqnarray}
Note that since $\omega_{cdm}$ contributes to the expansion rate before and after recombination it causes $k_D(z_*)$ to increase and $D_A(z_*)$ to decrease, leading to a small overall effect on $\theta_D$. Given the relatively small uncertainty in $\omega_{cdm}$ when determined from these data sets it makes a negligible contribution to the variation of $\theta_D$. Combining these we find that the small scale data gives a negative correlation between $n_s$ and $\omega_b$
\begin{equation}
\frac{\delta n_s}{n_s} \simeq -\frac{\delta \omega_b}{\omega_b}. \label{eq:small3}
\end{equation}

This indicates that on its own, a measurement of $\theta_D$ is not sufficient to break the degeneracy between $n_s$ and $\omega_b$. 
However, this degeneracy can be broken by adding information from intermediate scales. By requiring that the ratio of the heights of the first ($\mathcal{H}_1$ at $\ell_1 \simeq 215$) and second acoustic peak ($\mathcal{H}_2$ at $\ell_2 \simeq 530$) in the temperature power spectrum remain unchanged, one can derive
\begin{eqnarray}
      \delta\frac{\mathcal{H}_1}{\mathcal{H}_2} &\simeq& -2 \frac{\delta n_s}{n_s} + 1.4 \frac{\delta \omega_b}{\omega_b}- 0.09 \frac{\delta \omega_{cdm}}{\omega_{cdm}}, \nonumber \\
      &\xrightarrow[]{\delta \frac{\mathcal{H}_1}{\mathcal{H}_2}=0}& \frac{\delta n_s}{n_s} \simeq 0.7 \frac{\delta \omega_b}{{\omega_b}}- 0.045 \frac{\delta \omega_{cdm}}{\omega_{cdm}}\,.\label{eq:intermediate}
\end{eqnarray}
As in Eq.~(\ref{eq:small2}) the contribution from variations in the CDM physical density is typically negligible. 
When using only intermediate data, the parameter dependence of $\theta_D$ in Eq.~(\ref{eq:small2}) combined with Eq.~(\ref{eq:intermediate}) gives 
\begin{equation}
    \frac{\delta \theta_D}{\theta_D} \simeq -0.3 \frac{\delta n_s}{n_s}.\label{eq:intermediate2}
\end{equation}

\begin{figure}
    \centering
    \includegraphics[width=\columnwidth]{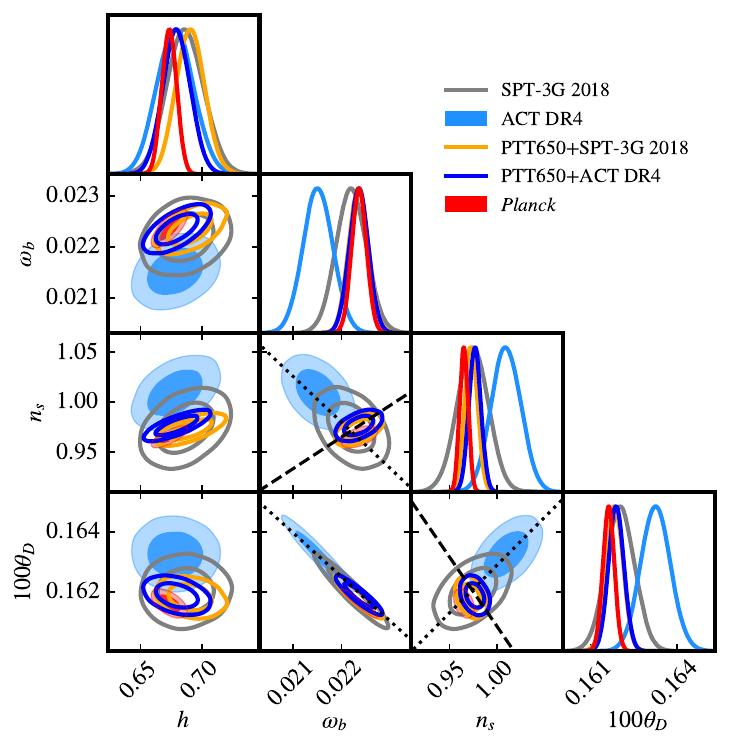}
    \caption{The triangle plot showing the 1D and 2D posterior distributions when fitting a variety of CMB data to $\Lambda$CDM. The dashed black lines correspond to the scaling Eqns.~(\ref{eq:intermediate}) and (\ref{eq:intermediate2}) and the dotted black lines correspond to the scaling in Eqns.~(\ref{eq:small1}), (\ref{eq:small2}), and (\ref{eq:small3}). }
    \label{fig:ACT_SPT_Planck_LCDM}
\end{figure}

These scaling relations allow us to see that the sign of the correlation between $n_s$ and $\omega_b$ changes when going from intermediate to small scales. This is confirmed by the dashed and dotted lines in Fig.~\ref{fig:ACT_SPT_Planck_LCDM}: SPT-3G 2018 and ACT DR4 mainly contain information from the damping tail and show a negative correlation between $n_s$ and $\omega_b$. However, once data sets that include intermediate scale information are considered (i.e., PTT650+SPT-3G 2018, PTT650+ACT DR4, and \textit{Planck}) the correlation flips to positive. These scaling relations allow us to accurately match the slope of the degeneracies, indicated by the black dashed and dotted lines. 

Fig.~\ref{fig:ACT_SPT_Planck_LCDM} makes it clear that ACT DR4 is in some tension with both \textit{Planck} and SPT-3G 2018 under $\Lambda$CDM. Several studies have found that \textit{Planck} and SPT-3G 2018 are statistically consistent, but inconsistent, at the $\sim 2-3 \sigma$ level, with ACT DR4 (see, e.g., Refs.~\cite{ACT:2020gnv,Handley:2020hdp}). The ACT collaboration has suggested that this may be due to an unexplained systematic error in the temperature/polarization calibration \cite{ACT:2020gnv} or due to physics beyond $\Lambda$CDM (see, e.g., Refs.~\cite{Hill:2021yec,Poulin:2021bjr,Smith:2022hwi}). 

As pointed out in Ref.~\cite{ACT:2020gnv}, one way to see the tension in the ACT DR4 data is in the $\omega_b-n_s$ plane. Unlike ACT DR4 (in light blue), the SPT-3G 2018 constraints (in gray) are in statistical agreement with \textit{Planck} (in red). When we add low to intermediate scale temperature data from \textit{Planck} to ACT DR4 (in dark blue)  and SPT-3G 2018 (in orange) the constraints considerably tighten, and both are in agreement with the full \textit{Planck} constraints. 

Another way to see the tension between ACT DR4 and \textit{Planck} is to compare their posteriors for $\theta_D$. We find that ACT DR4 gives $100\theta_D =0.16327\pm 0.00051$ and \textit{Planck} gives $100\theta_D = 0.16161\pm 0.00019$-- a tension of about 3.25$\sigma$. On the other hand SPT-3G 2018 is consistent with \textit{Planck} with $100 \theta_D = 0.16202 \pm 0.00051$. When PTT650 is combined with ACT DR4 we see that the posterior distribution for $\theta_D$ shifts to smaller values. Given that PTT650 does not directly measure $\theta_D$, this shift is caused by constraints placed on $\omega_b$ and $n_s$ which, in turn, pulls the value of $\theta_D$ down.

This discussion suggests that a cosmological model which introduces additional freedom in setting the damping scale may better accommodate the ACT DR4 preference for a higher $\theta_D$ (leading to higher $n_s$ and smaller $\omega_b$ under $\Lambda$CDM) while also providing an improved fit to the intermediate scales probed by PTT650. On the other hand, SPT-3G 2018 does not share this preference for a large $\theta_D$ indicating that it may not favor the same beyond $\Lambda$CDM physics as ACT DR4.

\subsection{Constraints on EDE}

\begin{figure*}[ht]
    \centering
    \ \includegraphics[width=\columnwidth]{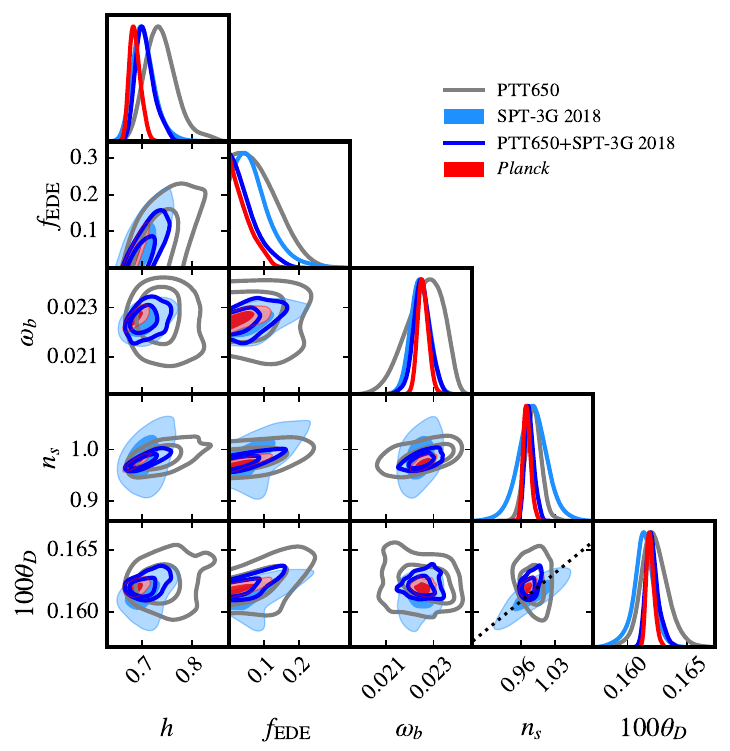}
    \includegraphics[width=\columnwidth]{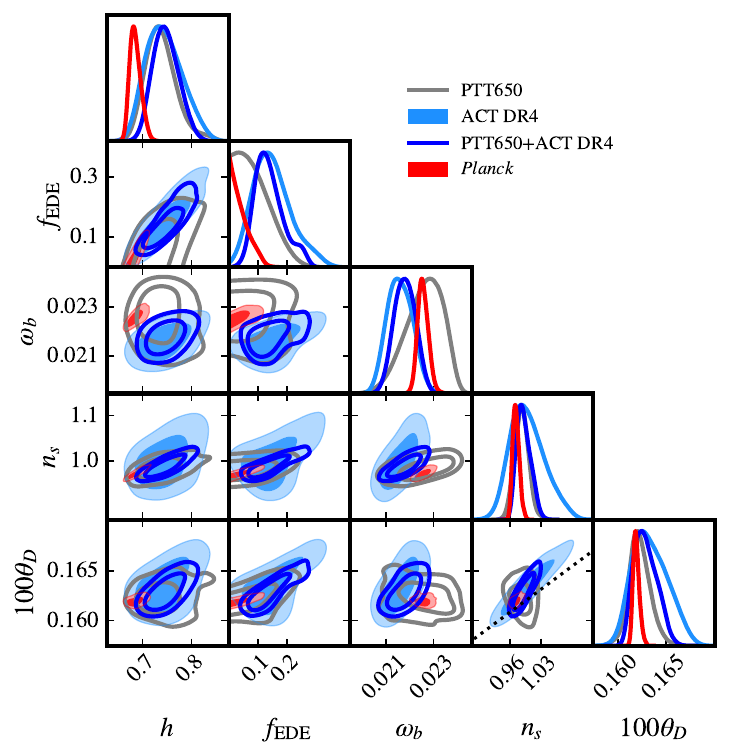}
    \caption{A triangle plot showing the 1D and 2D posterior distributions for EDE fits several different CMB data sets. The left panel shows fits including SPT-3G 2018 and the right panel shows fits including ACT DR4. The dotted line shows the expected degeneracy between $n_s$ and $\omega_b$ from small-scale CMB data in Eq.~(\ref{eq:small1}).}
    \label{fig:ACT_SPT_Planck_EDE}
\end{figure*} 

\begin{table*}[t]
    \scalebox{0.71}{
    \begin{tabular}{|l|c|c|c|c|c|c|c|c|} 
        \hline 
        Data & \multicolumn{2}{|c|}{SPT-3G 2018}& \multicolumn{2}{|c|}{PTT650+SPT-3G 2018} 
         &\multicolumn{2}{|c|}{ACT DR4} &\multicolumn{2}{|c|}{PTT650+ACT DR4}\\ \hline
        Model & $\Lambda$CDM & EDE & $\Lambda$CDM & EDE & $\Lambda$CDM & EDE & $\Lambda$CDM & EDE\\
        \hline
        \hline
        $f_{\rm EDE}$ & $-$ & $<0.172$ & $-$ & $<0.127$& $-$ & $0.154^{+0.049}_{-0.083}$ & $-$ & $0.138^{+0.033}_{-0.059}$ \\
        $\log_{10}z_c$ & $-$ & unconstrained
        &$-$ & unconstrained
        &$-$ & $< 3.76$
        &$-$ & $3.27^{+0.10}_{-0.12}$\\
        $\theta_i$ &  $-$ & unconstrained
        &  $-$ & unconstrained
        &  $-$ & unconstrained
        &  $-$ & unconstrained\\
        \hline
        $h$ &$0.686\pm 0.015$  & $0.709^{+0.018}_{-0.022}$ &$0.690\pm 0.012$  & $0.705^{+0.014}_{-0.020}$
        &$0.678^{+0.014}_{-0.016}$  & $0.745^{+0.032}_{-0.044}$ &$0.680\pm 0.012$  & $0.746^{+0.024}_{-0.029}$\\
        $\omega_b$ & $0.02220\pm 0.00031$  & $0.02253\pm 0.00034$
        & $0.02236\pm 0.00020$  & $0.02248\pm 0.00037$
        & $0.02151\pm 0.00030$  & $0.02159\pm 0.00054$
        & $0.02235\pm 0.00021$  & $0.02175\pm 0.00045$\\
        $\omega_{\rm cdm}$ &$0.1165\pm 0.0038$ & $0.1243^{+0.0050}_{-0.0063}$
         &$0.1158\pm 0.0028$ & $0.1207^{+0.0033}_{-0.0054}$
         &$0.1182\pm 0.0037$ & $0.1353^{+0.0078}_{-0.013}$
         &$0.1196\pm 0.0029$ & $0.1325^{+0.0063}_{-0.010}$\\
        $10^{9}A_s$ & $2.079\pm 0.042$ & $2.076\pm 0.046$
        & $2.070\pm 0.034$ & $2.085\pm 0.039$
        & $2.072\pm 0.040$ & $2.127^{+0.051}_{-0.063}$
        & $2.114\pm 0.034$ & $2.128\pm 0.056$\\
        $n_s$&  $0.975\pm 0.016$ &  $1.002^{+0.023}_{-0.025}$
        &  $0.9727\pm 0.0066$ &  $0.9772^{+0.0080}_{-0.0094}$
        &  $1.010\pm 0.015$ &  $1.002^{+0.033}_{-0.048}$
        &  $0.9768\pm 0.0068$ &  $0.989^{+0.013}_{-0.019}$\\
        \hline
        $\sigma_8$ &$0.800\pm 0.015$ & $0.816\pm 0.018$
        &$0.795\pm 0.013$ & $0.806\pm 0.018$
        &$0.820^{+0.015}_{-0.013}$ & $0.844\pm 0.036$
        &$0.819\pm 0.013$ & $0.837\pm 0.031$\\
        $\Omega_m$ &$0.297^{+0.019}_{-0.022}$ & $0.294^{+0.017}_{-0.021}$
        &$0.292\pm 0.015$ & $0.290\pm 0.018$
        &$0.306\pm 0.021$ & $0.285^{+0.021}_{-0.023}$
        &$0.309\pm 0.017$ & $0.279\pm 0.017$\\
        Age [Gyrs] & $13.787\pm 0.046$&$13.38^{+0.28}_{-0.16}$
        & $13.763\pm 0.038$&$13.51^{+0.22}_{-0.12}$
        & $13.830^{+0.049}_{-0.043}$&$12.87^{+0.58}_{-0.38}$
        & $13.752\pm 0.041$&$12.91^{+0.42}_{-0.30}$\\
        100$\theta_s$ & $1.04203\pm 0.00074$&$1.04119^{+0.00091}_{-0.00082}$
        & $1.04218\pm 0.00065$&$1.04174^{+0.00075}_{-0.00066}$
        & $1.04338\pm 0.00071$&$1.04243\pm 0.00081$
        & $1.04319\pm 0.00067$&$1.04231\pm 0.00067$\\
        100$\theta_D$ & $0.16202\pm 0.00051$&$0.16281^{+0.00062}_{-0.00074}$
        & $0.16182\pm 0.00029$&$0.16203^{+0.00042}_{-0.00051}$
        & $0.16327\pm 0.00051$&$0.1635^{+0.0017}_{-0.0023}$
        & $0.16190\pm 0.00028$&$0.16280^{+0.00093}_{-0.0015}$\\
        \hline
        $\Delta \chi^2_{\rm min}$ (EDE$-\Lambda$CDM) & $-$ & -2.7 & $-$ & -1.1 & $-$ & -9.4 & $-$ &-15.9\\
        \hline
    \end{tabular} 
    }
    \caption{The mean $\pm 1\sigma$ uncertainties of the cosmological parameters for the SPT-3G 2018 and ACT DR4 data sets. All limits are at the 95\% confidence level.}
    \label{tab:full}
\end{table*}

Any cosmological model that introduces additional energy density solely before recombination\footnote{In the case of the EDE model we are considering here, this is true as long as $\log_{10}z_c \gtrsim 3.3$.} with fixed $\theta_s$ generically predicts an increase in $\theta_D$ \cite{Poulin:2023lkg}, therefore opening the possibility of constraining a generic EDE resolution of the Hubble tension with high angular resolution measurements, such as those from ACT DR4 and SPT-3G 2018.

In Fig.~\ref{fig:ACT_SPT_Planck_EDE} we show the 2D posterior distributions of $\{h,f_{\rm EDE},\omega_b,n_s,100\theta_D\}$ when analyzing SPT-3G 2018 (left panel) or ACT DR4 (right panel), alone or in combination with PTT650. We compare these posteriors to those obtained when analyzing {\it Planck} and the results of these MCMC analyses are reported in Table \ref{tab:full}. A triangle plot comparing all cosmological parameters reconstructed from the three experiments is provided in Fig.~\ref{fig:big2D} in the Appendix.

There is a stark difference between the results of analyses of SPT-3G 2018 and ACT DR4. As shown in the left panel of Fig.~\ref{fig:ACT_SPT_Planck_EDE}, SPT-3G 2018 data alone do not favor EDE
and the combination of PTT650 and SPT-3G 2018 provides upper limits on $f_{\rm EDE} <0.127$ that are in agreement (albeit weaker) with the full \textit{Planck} data set, $f_{\rm EDE}<0.091$ \cite{Hill:2020osr,Simon:2022adh}. This is in contrast with the ACT DR4 data, shown in the right panel, which shows a $2-3\sigma$ preference for $f_{\rm EDE}>0$ with or without PTT650 as reported previously \cite{Hill:2021yec,Poulin:2021bjr,Smith:2022hwi}. 

The constraints to EDE using SPT-3G 2018 (light blue) show a positive correlation between $n_s$ and $\theta_D$, with a slope which is consistent with keeping the amplitude of the small-scale power spectrum fixed (i.e., Eq.~(\ref{eq:small1}), shown by the dotted line). The PTT650 constraints (gray) show no correlation between $n_s$ and $\theta_D$. We can also see that the parameter degeneracy between $n_s$ and $\omega_b$ for SPT-3G 2018 and PTT650 are nearly orthogonal. The resulting joint constraints tighten the posterior distributions for $\omega_b$, $n_s$, and $\theta_D$, and the positive correlation between $f_{\rm EDE}$ and $\theta_D$ leads to a tighter upper limit on $f_{\rm EDE}$. It is also interesting to note that the SPT-3G 2018 upper limit on $\theta_D$ remains unchanged when we add PTT650, indicating that even in the joint constraints the angular damping scale is being constrained by the small-scale measurements.

In the case of ACT DR4, on the other hand, one can see that the degeneracy between $100\theta_D$ and $f_{\rm EDE}$ is much more pronounced, leading to wider posterior distributions for $\theta_D$ and $n_s$.  This improves the overlap with {\it Planck}, and explains why, once PTT650 is added, the preference for EDE further increases. However, note that the strong negative correlation between $\theta_D$ and $\omega_b$ in Eq.~(\ref{eq:small2}) is absent when fit to EDE. As a result, the preference for a lower $\omega_b$ seen in ACT DR4 persists despite the presence of EDE and broader $\theta_D$. This leads to a small cost in the fit to the PTT650 data, $(\chi^2_{\rm PTT650})_{\rm EDE}-(\chi^2_{\rm PTT650})_{\Lambda{\rm CDM}} = 0.59$ with $f_{\rm EDE} = 0.11$ and $h=0.737$ compared to $h=0.675$. We also note that, unlike for SPT-3G 2018, the upper limit to $\theta_D$ changes significantly when we add PTT650 to ACT DR4. This indicates that the joint constraints are not directly probing the angular damping scale, but instead the upper limit on $\theta_D$ is driven by constraints on the parameters it depends on. 

\begin{figure}
    \centering
    \includegraphics[width=\columnwidth]{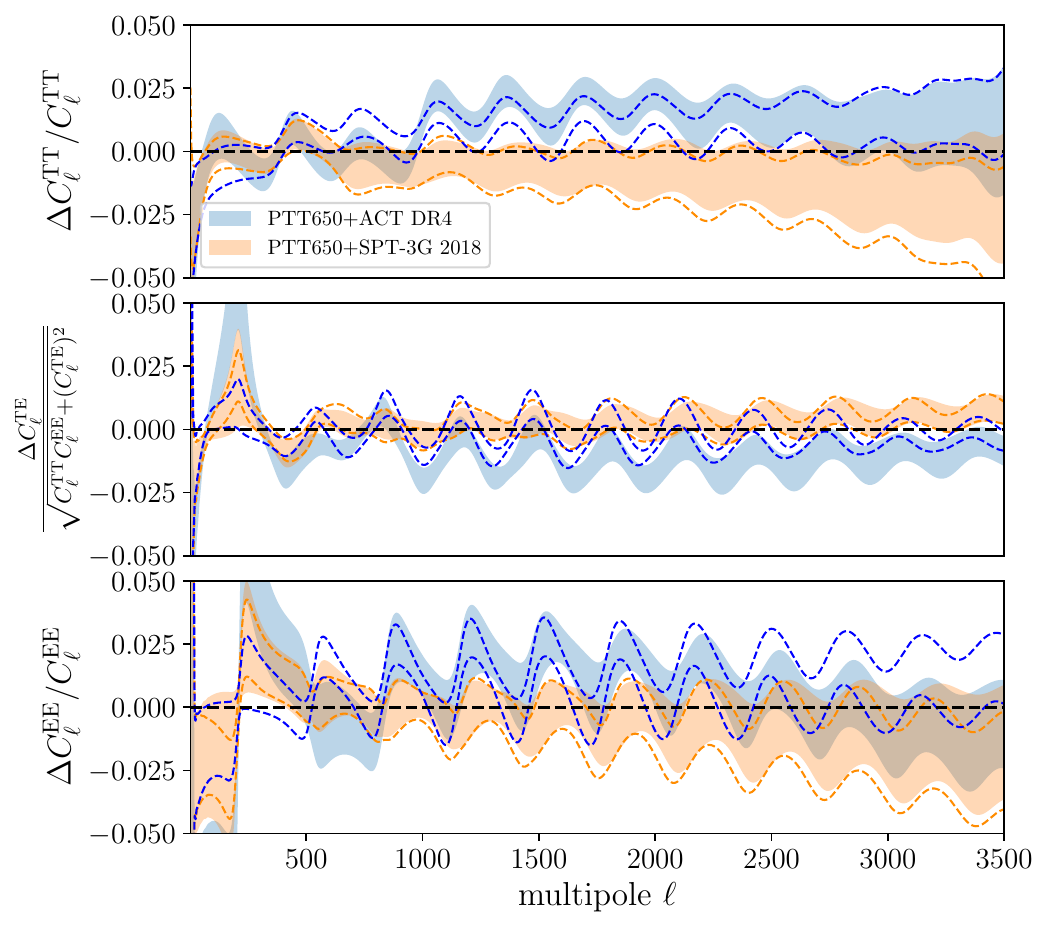}
    \caption{The power spectrum residuals (with respect to the \textit{Planck} 2018 best fit $\Lambda$CDM power spectra) for PTT650+ACT DR4 and PTT650+SPT-3G 2018 fit to EDE (filled bands) and $\Lambda$CDM (dashed lines). The bands were generated by drawing samples from the MCMC chains and computing the 68\% confidence interval at each multipole.}
    \label{fig:ACT_SPT_residuals}
\end{figure}

To understand the difference between ACT DR4 and SPT-3G 2018, it is instructive to look at a comparison between their residuals. Fig.~\ref{fig:ACT_SPT_residuals} shows the 68\% CL region of the residuals at each multipole, $\ell$, computed from 100 random samples from the MCMC posteriors in both EDE (filled bands) and $\Lambda$CDM (dashed lines), taken with respect to the corresponding \textit{Planck} 2018 best fit $\Lambda$CDM power spectra.
It is striking that the residuals are noticeably different between SPT-3G 2018 and ACT DR4 (in both EDE and $\Lambda$CDM), which is illustrating some level of inconsistency between the two data sets. 

For SPT-3G 2018, there is essentially no difference in the residuals when fit to EDE or $\Lambda$CDM, confirming the fact that the SPT-3G 2018 data do not favor EDE over \LCDM{}. They show a mild decrement at the higher multipoles in TT and EE and are compatible with zero at all multipoles. For ACT DR4, the $\Lambda$CDM and EDE residuals also have a qualitatively similar shape in TT and EE, displaying a characteristic `step' around $\ell \simeq 1500$ to an enhancement of power, with only small differences in TT and EE at intermediate multipoles ($\ell \sim 500$). The most notable difference is in the temperature/E-mode cross power spectrum (TE) residuals, that oscillate around zero in $\Lambda$CDM but are offset from zero in EDE. This agrees with Ref.~\cite{Hill:2021yec} which found that for this data combination the TE spectrum is the main driver of the preference for EDE.

These residuals can be understood in light of the parameter constraints, although it can appear counter-intuitive: at the parameter level the ACT DR4 fit prefers a larger value of $\theta_D$ which leads to a \emph{suppression} of power on small scales. This seems to contradict the enhanced power we see in Fig.~\ref{fig:ACT_SPT_residuals}.However, as listed in Table \ref{tab:full}, the PTT650+ACT DR4 mean values for $A_s$ and $n_s$ are larger than those for the $\Lambda$CDM best fit to \textit{Planck} ($A_s^{\Lambda{\rm CDM}}= 2.10058 \times 10^{-9}$ and $n_s^{\Lambda{\rm CDM}} =  0.96605$): $\Delta A_s/\sigma_{A_s} \simeq 0.4$ and $\Delta n_s/\sigma_{n_s} \simeq 1.6$ for $\Lambda$CDM and $\Delta A_s/\sigma_{A_s} \simeq 0.5$ and $\Delta n_s/\sigma_{n_s} \simeq 1.2$ for EDE. The increase in the small-scale amplitude due to these shifts is counteracted by the increased damping from the increase in $\theta_D$, leading to the residual excess of about 2\% seen in Fig.~\ref{fig:ACT_SPT_residuals}. On the other hand the reduction in power for the PTT650+SPT-3G 2018 residuals is explained by an increase in $\theta_D$ relative to the $\Lambda$CDM \textit{Planck} best fit value ($\theta^{\Lambda{\rm CDM}}_D = 0.16139$): $\Delta \theta_D/\sigma_{\theta_D} = 1.5$ for $\Lambda$CDM and $\Delta \theta_D/\sigma_{\theta_D} = 1.25$ for EDE.

In order to estimate the extent to which ACT DR4 and SPT-3G 2018 are statistically compatible, we make use of the {\sf Tensiometer} package\footnote{\url{https://github.com/mraveri/tensiometer}} \cite{Raveri:2021wfz} and compute the `parameter shift' tension between these two datasets in both EDE and $\Lambda$CDM. In the case of $\Lambda$CDM the disagremeent is at the $1.7\sigma$ level, and increases to the $2.9\sigma$ level in EDE. Although the tension remains at a statistically `acceptable' level (i.e., one could argue that they are statistical fluctuations), future measurements of the CMB damping tail will be important to assess this inconsistency, and the true level of constraints on EDE. 
 
\subsection{EDE constraints using TT vs.~TE/EE}

Given the results in the previous subsection it is of interest to further explore what drives the constraints to EDE by considering how the model fits different subsets of the data. One natural way to do this is to look at constraints from temperature and polarization power spectra separately. 

\begin{figure*}
    \centering
    \includegraphics[width=\columnwidth]{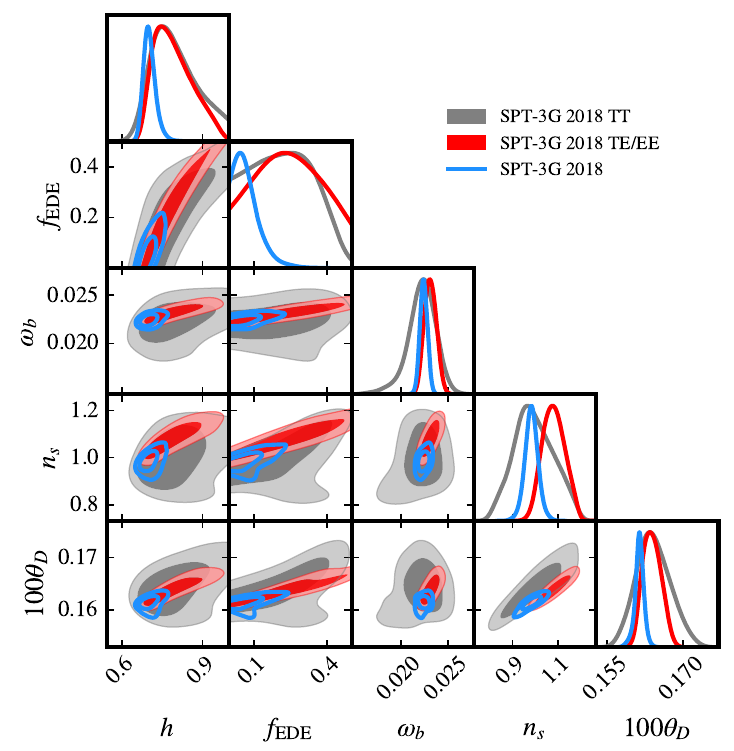}
    \includegraphics[width=\columnwidth]{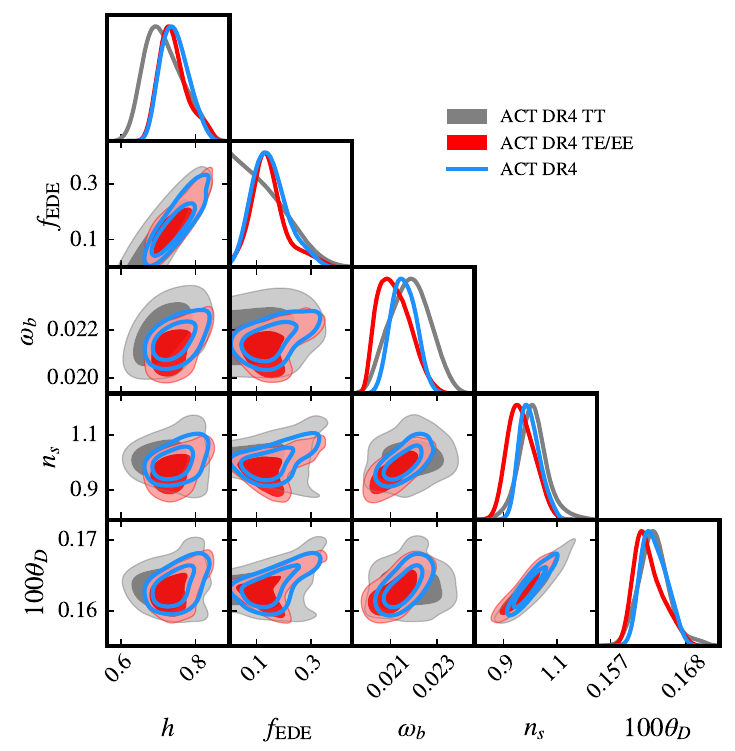}
    \caption{A triangle plot showing the posterior distributions for EDE fits to SPT-3G 2018 temperature and polarization data, separately.}
    \label{fig:SPT_TT_vs_TEEE}
\end{figure*}

The division of the data into temperature and polarization provides insights into these constraints for several reasons. First it has been established that the different physical origins for temperature and polarization perturbations imply that they will produce different degeneracies between cosmological parameters (see, e.g., Refs.~\cite{Eisenstein:1998hr,Colombo:2008ta,Galli:2010it,Wu:2014hta}). In addition to this, several studies have pointed out that assuming the same noise levels, CMB polarization better constrains cosmology than temperature \cite{Rocha:2003gc,Galli:2014kla}. It is well known that at small angular scales the astrophysical foregrounds are expected to have a reduced impact on polarization compared to temperature (see, e.g., Ref.~\cite{Tucci_2012}), so we expect such a split to have potentially significantly different systematic errors. Finally, it is of practical use since it allows us to compare what we find here to previous analyses of SPT-3G 2018 data on EDE which have only had access to polarization information.

\begin{figure}
    \centering
    \includegraphics[width=\columnwidth]{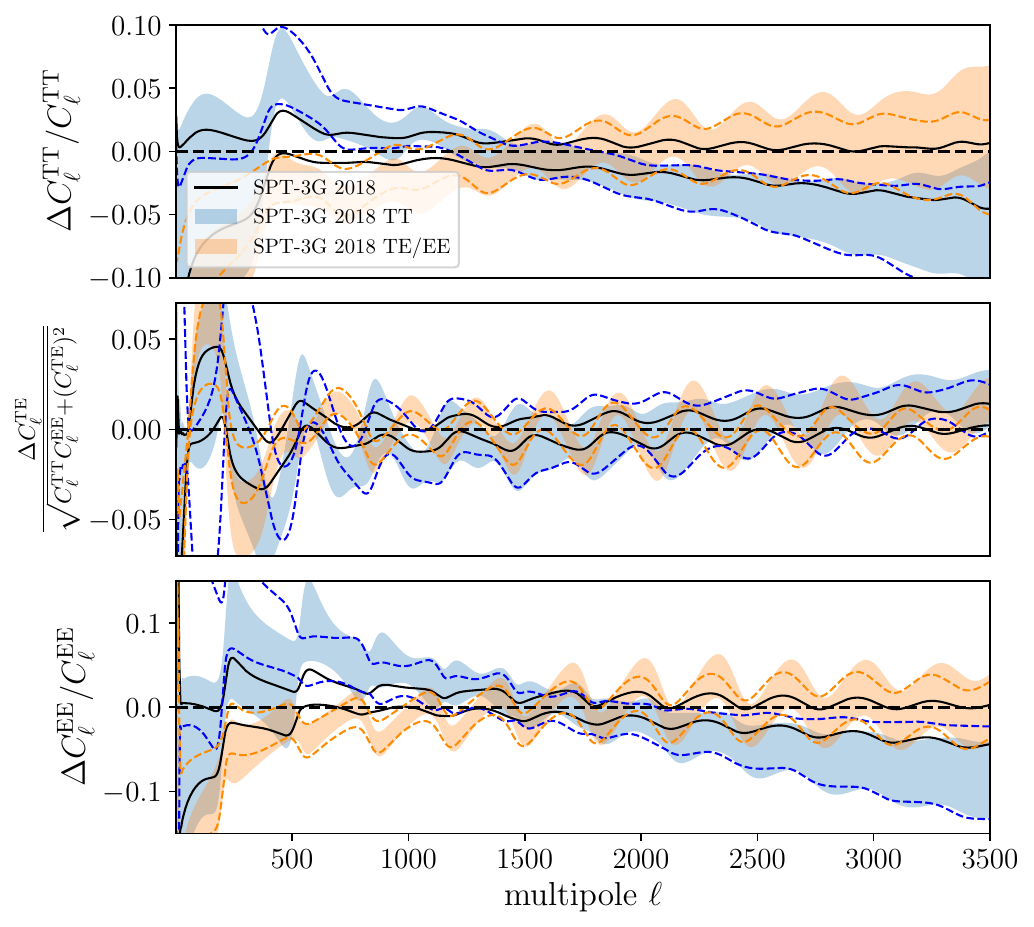}
    \caption{The SPT-3G 2018 fractional residuals with respect to the \textit{Planck} best fit $\Lambda$CDM model \cite{Planck:2018vyg}. The dashed lines show residuals from $\Lambda$CDM and the filled regions show residuals from EDE. The residuals were generated by drawing samples from the MCMC chains and computing the 68\% confidence interval at each multipole.}
    \label{fig:SPT_residual}
\end{figure}

The results of this analysis for SPT-3G 2018 and ACT DR4 are shown in Fig.~\ref{fig:SPT_TT_vs_TEEE}. The SPT-3G 2018 constraints in the left panel shows some `curious' results. First, the temperature and polarization measurements are, separately, consistent with large values of $f_{\rm EDE}$ and correspondingly large values of $h = 0.8 \pm 0.1$. However, when the TT/TE/EE data set is used, one finds that the uncertainty on both parameters is significantly smaller, with $f_{\rm EDE} = 0.089^{+0.037}_{-0.053}$ and $h=0.709^{+0.018}_{-0.022}$. This is reminiscent of what happens for \textit{Planck}, where TT and TE/EE constraints are weaker than the TT/TE/EE data set \cite{Smith:2022hwi,Poulin:2023lkg}. On the other hand, the ACT DR4 constraints in the right panel show that both temperature and polarization posteriors are similar to those using the TT/TE/EE data set. 

The increase in sensitivity to $f_{\rm EDE}$ when using both SPT-3G 2018 temperature and polarization does not appear to come from a simple parameter degeneracy. The only parameter with a slightly discrepant posterior distribution is $n_s$, with polarization preferring a slightly larger value than the temperature measurements. Looking at the 2D posterior distribution in the $n_s$-$f_{\rm EDE}$ plane in the left panel of Fig.~\ref{fig:SPT_TT_vs_TEEE} we can see that the overlap between the 1$\sigma$ TT (gray) and TE/EE (red) contours is in fact larger for large values of $f_{\rm EDE}$, and includes parameter space where $f_{\rm EDE}$ can be as large as 0.4, indicating that the SPT-3G 2018 constraint on $f_{\rm EDE}$ cannot be simply described through differences in their constraints on $n_s$.

Going beyond a comparison between parameters, we plot the residuals in Fig.~\ref{fig:SPT_residual} with respect to the $\Lambda$CDM bestfit to {\it Planck} data. We show the EDE residuals with filled bands and the $\Lambda$CDM ones with dashed lines. There it is clear that when using SPT-3G 2018 temperature measurements (blue band) the residuals prefer to have excess/deficit in power at larger/smaller scales, whereas the polarization prefers the opposite, in both EDE and $\Lambda$CDM. The residuals for the total data set split the difference, leading to significantly tighter constraints than each part separately. We note that changes to $n_s$ would induce a tilt centered around $l_p\simeq 550$ (which corresponds to a pivot wavenumber $k_p = 0.05 \ {\rm Mpc}^{-1}$). This scale is significantly lower than the scale at which the SPT-3G 2018 TT vs.~TE/EE residuals cross, $l\simeq 1500$, providing further evidence that the difference in the TT vs.~TE/EE constraints is not simply driven by shifts in $n_s$.

Fig.~\ref{fig:SPT_residual} suggests that there is some tension between the temperature and polarization residuals.  Although it is beyond the scope of this work to determine the level of tension in the residuals/spectra, we have used {\sf Tensiometer} to estimate the `parameter shift' tension between SPT-3G 2018 TT and TE/EE: when fitting $\Lambda$CDM we find a good agreement at the 1$\sigma$ level despite the apparent discrepancy seen in the shape of the residuals, while when fitting EDE we find a disagreement at the 2.3$\sigma$ level. For comparison, the same analysis applied to the \textit{Planck} TT and TE/EE power spectra gives agreement at the 0.3$\sigma$ level in $\Lambda$CDM but disagreement at the 2.7$\sigma$ level in EDE (see Ref.~\cite{Smith:2022hwi} for a discussion around potential systematic effects in TE/EE with a focus on EDE). Finally, we find in the case of ACT DR4 that the TT and TE/EE data are in agreement at the 0.4$\sigma$ level ($\Lambda$CDM) and 0.1$\sigma$ level (EDE). 

A similar result was reported in Ref.~\cite{SPT-3G:2022hvq} when quoting constraints on primordial magnetic fields. The presence of primordial magnetic fields causes a boost in the baryon density perturbations which, in turn, induces additional fluctuations in the CMB temperature and polarization. The constraints to the amplitude of this boost, $b$, are weak when using SPT-3G 2018 TT or TE/EE but significantly strengthen when using TT/TE/EE (see Figs.~9 and 12 of Ref.~\cite{SPT-3G:2022hvq}). Ref.~\cite{SPT-3G:2022hvq} investigated this by generating mock SPT-3G 2018 bandpowers using the measured covariance matrix and found that the limits to $b$ were within 20\% of the expected constraints assuming $b=0$. The similarity of the results presented here and in Ref.~\cite{SPT-3G:2022hvq} points to the conclusion that the SPT-3G 2018 constraints on EDE are statistically consistent. However, to be certain of this, one would have to perform a similar mock analysis to further assess the statistical consistency of the SPT-3G 2018 constraints on EDE. We leave such an in-depth analysis of the differences between the SPT-3G 2018 temperature and polarization measurements to future work. 

\section{The residual tension with S$H_0$ES}
\label{sec:ext}

We now turn to combining CMB observations with other cosmological data sets, to compute the strongest constraints to EDE to date, and gauge the residual level of tension with S$H_0$ES. To mitigate prior volume effects (see Refs.~\cite{Smith:2019ihp,Smith:2020rxx,Herold:2022iib,Herold:2021ksg} for further discussion), we compute the tension metric $Q_{\rm DMAP}\equiv\sqrt{\Delta\chi^2({\rm w/~S}H_0{\rm ES})-\Delta\chi^2({\rm w/o~S}H_0{\rm ES})}$ \cite{Raveri:2018wln} rather than assuming Gaussian posterior distributions. We perform analyses of \Planck~alone, \Planck+SPT-3G 2018, \Planck+SPT-3G 2018+ACT DR4, always including the CMB lensing, BAO, and Pantheon+ data sets (denoted as external data sets, `Ext') described in Sec.~\ref{sec:method}. Cosmological parameters credible intervals are reported in the Appendix (Tab.~\ref{tab:EDE_MCMC_withExt} and  $\chi^2$ statistics are provided in Tab.~\ref{tab:EDE_chi2_withExt}). 

\begin{figure}
    \centering
    \includegraphics[width=1\columnwidth]{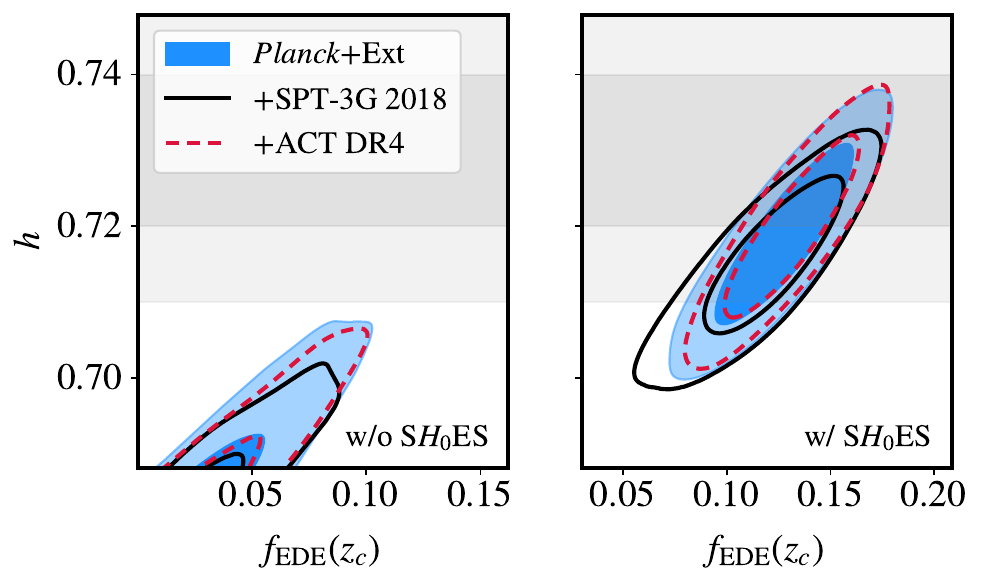}
    \caption{Posterior distribution of $h$ and $f_{\rm EDE}$ with (right panel) and without (left panel) the inclusion of the S$H_0$ES prior on $M_b$. The combination of \textit{Planck}+SPT-3G 2018 restricts the degeneracy between $h$ and $f_{\rm EDE}$ compared to using \textit{Planck} alone. The inclusion of ACT DR4 weakens the constraints to $f_{\rm EDE}$, allowing for a better fit of S$H_0$ES in the combined analysis.}
    \label{fig:EDE_SPT_ACT_Ext_Mb}
\end{figure}

Fig.~\ref{fig:EDE_SPT_ACT_Ext_Mb} shows the posterior distributions of $f_{\rm EDE}$ and $h$ when we combine CMB observations with the external cosmological data sets and with or without S$H_0$ES. When considering {\em Planck} EDE reduces the Hubble tension to $2.6\sigma$\footnote{This level of tension is higher than previously reported (i.e., 1.6$\sigma$ from Table 1 of Ref.~\cite{Schoneberg_2019}) due to the use of SNeIa data from Pantheon+ \cite{Brout:2022vxf} instead of Pantheon \cite{Pan-STARRS1:2017jku}}; when adding SPT-3G 2018 the tension goes up to $2.9\sigma$. When S$H_0$ES is left out of the analysis, we obtain a bound $f_{\rm EDE}<0.071$ (to be interpreted with some degree of caution given the known prior volume effects), while the inclusion of the S$H_0$ES prior leads to a $\gtrsim 5\sigma$ detection of $f_{\rm EDE}=0.121^{+0.024}_{-0.019}$. The inclusion of ACT DR4, which pulls the EDE contribution up along with an increase in $h$, reduces the tension to $1.6\sigma$, but the discrepancy between ACT DR4 and {\it Planck}+SPT-3G 2018 casts some doubts on the statistical consistency of this result.

Given that the SPT-3G 2018 is in good statistical agreement with \textit{Planck} and that the inclusion of SPT-3G 2018 increases the Hubble tension over using \textit{Planck} alone, it is clear that the TT/TE/EE SPT-3G 2018 data set provides evidence against the hint of EDE seen in ACT DR4. The next CMB data release by the ACT collaboration is eagerly awaited to shed light on this apparent inconsistency.

\section{Conclusions}
\label{sec:concl}

In this paper we have set constraints on the axion-like EDE model using the recently released temperature and polarization power spectra from the SPT-3G 2018 collaboration \cite{SPT-3G:2022hvq}. These are particularly important given the apparent disagreement between {\it Planck} and ACT DR4: while EDE only marginally improves the fit to {\it Planck} over $\Lambda$CDM, with no detection of EDE in a Bayesian analysis, ACT DR4 favors a non-zero EDE contribution at the $2-3\sigma$ level. These results were shown to originate from some apparent (statistically mild) inconsistency between ACT DR4 and {\it Planck}, in particular at high-$\ell$ in temperature (on top of some differences in polarization at intermediate multipoles). The new temperature and polarization measurements from SPT-3G 2018 therefore have the ability to arbitrate the difference between ACT DR4 and {\it Planck}. We have found that SPT-3G 2018 on its own does not favor EDE, and places a weak constraint of $f_{\rm EDE}<0.172$. When combined with PTT650, they become nearly as constraining as the full \textit{Planck} data set, and disfavor the cosmological origin of the signal seen in ACT DR4.

At least some of the constraining power from SPT-3G 2018 comes from its limits on the angular damping scale, $\theta_D$, and in turn to the constraints put on $n_s$ and $\omega_b$, highlighting that $\theta_D$ measured with ACT DR4 differs at the $2-3\sigma$ level from that measured with {\it Planck} and SPT-3G 2018 (which are in good agreement with each other). This translates into preference for a larger value of $n_s$ and a smaller value of $\omega_b$ under $\Lambda$CDM within ACT DR4. When EDE is included, the posterior of $\theta_D$  and $n_s$ becomes wider in ACT DR4, improving the overlap with other CMB experiments, and driving the preference for EDE. However, $\omega_b$ remains lower than in {\it Planck} and SPT-3G 2018, driving $f_{\rm EDE}$ to zero in a combined analyses of all three experiments. 

We also show that there is some `curiosity' when looking at EDE fits to SPT-3G 2018 TT and TE/EE separately. The combined analysis places significantly tighter constraints on EDE than either of them individually, with the individual constraints saturating our prior on $f_{\rm EDE} < 0.5$, but the TT/TE/EE SPT-3G 2018 data set gives $f_{\rm EDE}<0.172$. This significant increase in sensitivity to $f_{\rm EDE}$ is not reflected at the level of the parameter posterior distributions. A similar result was found in Ref.~\cite{SPT-3G:2022hvq} when constraining the presence of primordial magnetic fields. A simulated band-power analysis showed that the actual SPT-3G 2018 constraints were within 20\% of the simulated ones, indicating that the constraints are statistically consistent. Given the similarity to what we have found with EDE, it is likely that the constraints presented here are similarly statistically consistent.

Looking at the power spectra residuals, Fig.~\ref{fig:SPT_TT_vs_TEEE} shows that the fit to SPT-3G 2018 TT produces residuals which have excess power at larger scales and a deficit of power at smaller scales; the opposite is true of the EE residuals. The combination of the two produces posterior distributions and residuals which are much more constrained than either individually.  We leave it to future work to conclusively determine whether the residuals when fit to TT and EE are consistent with expected statistical fluctuations. 

Finally, we have established that the ability for EDE to resolve the Hubble tension is reduced when SPT-3G 2018 data are included. We quantify the reduction in the tension between the CMB and the S$H_0$ES data by computing the $Q_{\rm DMAP}$ tension metric \cite{Raveri:2018wln,Schoneberg:2021qvd} and find that the tension goes up from $2.6\sigma$ (with {\it Planck} alone) to $2.9\sigma$ (with {\it Planck} + SPT). The inclusion of ACT DR4 reduces the tension to $1.6\sigma$ since ACT DR4 favors larger EDE fraction, with the caveat that ACT DR4 is the outlier.  Although we have not performed a profile likelihood analyses, the degradation in the $Q_{\rm DMAP}$ metric indicates that the additional constraining power from SPT is not solely driven by prior volume effects. 

Looking towards the near future we expect to have new data releases from both the SPT and ACT collaborations as well as data from the Simons Observatory (currently under construction) and CMB-S4 (currently in an advanced planning stage). All of these ground-based CMB telescopes complement what has already been measured from space by \textit{Planck} by providing us with independent measurements at intermediate angular scales and extending measurements to smaller scales. Previous work has emphasized how the new small-scale measurements may uniquely probe the impact of EDE through better constraints on the shape of the damping tail (i.e., Ref.~\cite{Poulin:2023lkg}). The results we have presented here indicate that sensitivity to EDE will come from a combination of both intermediate and small-scale measurements, in order to break parameter degeneracies, as well as from the complementarity between temperature and polarization power spectra. These results help to better focus model building efforts in order to develop theories which can successfully address the Hubble tension. 

\begin{acknowledgements}

The authors thank Lennart Balkenhol for extensive help with the SPT-3G 2018 TT/TE/EE likelihood, Karim Benabed for providing the authors with a {\sf MontePython}-compatible version of the SPT-3G 2018 TT/TE/EE likelihood, Silvia Galli and Ali Rida Khalife for conversations about fitting beyond-\LCDM{} cosmologies to the SPT-3G 2018 data, and Marc Kamionkowski and Adam Riess for conversations about the current state of the Hubble tension. TLS is supported by NSF Grant No.~2009377. This work used the Strelka Computing Cluster, which is supported by the Swarthmore College Office of the Provost.  VP is supported by funding from the European Research Council (ERC) under the European Union’s HORIZON-ERC-2022 (Grant agreement No.~101076865).  The authors acknowledge the use of computational resources from the Excellence Initiative of Aix-Marseille University (A*MIDEX) of the “Investissements d’Avenir” programme. These results have also been made possible thanks to LUPM's cloud computing infrastructure founded by Ocevu labex, and France-Grilles. This project has received support from the European Union’s Horizon 2020 research and innovation program under the Marie Skodowska-Curie grant agreement No.~860881-HIDDeN.

\end{acknowledgements}

\bibliography{biblio.bib}

\begin{thebibliography}{55}%
\makeatletter
\providecommand \@ifxundefined [1]{%
 \@ifx{#1\undefined}
}%
\providecommand \@ifnum [1]{%
 \ifnum #1\expandafter \@firstoftwo
 \else \expandafter \@secondoftwo
 \fi
}%
\providecommand \@ifx [1]{%
 \ifx #1\expandafter \@firstoftwo
 \else \expandafter \@secondoftwo
 \fi
}%
\providecommand \natexlab [1]{#1}%
\providecommand \enquote  [1]{``#1''}%
\providecommand \bibnamefont  [1]{#1}%
\providecommand \bibfnamefont [1]{#1}%
\providecommand \citenamefont [1]{#1}%
\providecommand \href@noop [0]{\@secondoftwo}%
\providecommand \href [0]{\begingroup \@sanitize@url \@href}%
\providecommand \@href[1]{\@@startlink{#1}\@@href}%
\providecommand \@@href[1]{\endgroup#1\@@endlink}%
\providecommand \@sanitize@url [0]{\catcode `\\12\catcode `\$12\catcode
  `\&12\catcode `\#12\catcode `\^12\catcode `\_12\catcode `\%12\relax}%
\providecommand \@@startlink[1]{}%
\providecommand \@@endlink[0]{}%
\providecommand \url  [0]{\begingroup\@sanitize@url \@url }%
\providecommand \@url [1]{\endgroup\@href {#1}{\urlprefix }}%
\providecommand \urlprefix  [0]{URL }%
\providecommand \Eprint [0]{\href }%
\providecommand \doibase [0]{http://dx.doi.org/}%
\providecommand \selectlanguage [0]{\@gobble}%
\providecommand \bibinfo  [0]{\@secondoftwo}%
\providecommand \bibfield  [0]{\@secondoftwo}%
\providecommand \translation [1]{[#1]}%
\providecommand \BibitemOpen [0]{}%
\providecommand \bibitemStop [0]{}%
\providecommand \bibitemNoStop [0]{.\EOS\space}%
\providecommand \EOS [0]{\spacefactor3000\relax}%
\providecommand \BibitemShut  [1]{\csname bibitem#1\endcsname}%
\let\auto@bib@innerbib\@empty
\bibitem [{\citenamefont {Turner}(2022)}]{Turner:2022gvw}%
  \BibitemOpen
  \bibfield  {author} {\bibinfo {author} {\bibfnamefont {Michael~S.}\
  \bibnamefont {Turner}},\ }\bibfield  {title} {\enquote {\bibinfo {title}
  {{The Road to Precision Cosmology}},}\ }\href {\doibase
  10.1146/annurev-nucl-111119-041046} {\  (\bibinfo {year} {2022}),\
  10.1146/annurev-nucl-111119-041046},\ \Eprint
  {http://arxiv.org/abs/2201.04741} {arXiv:2201.04741 [astro-ph.CO]}
  \BibitemShut {NoStop}%
\bibitem [{\citenamefont {{Peebles}}(2020)}]{2020RvMP...92c0501P}%
  \BibitemOpen
  \bibfield  {author} {\bibinfo {author} {\bibfnamefont {P.~J.~E.}\
  \bibnamefont {{Peebles}}},\ }\bibfield  {title} {\enquote {\bibinfo {title}
  {{Nobel Lecture: How physical cosmology grew*}},}\ }\href {\doibase
  10.1103/RevModPhys.92.030501} {\bibfield  {journal} {\bibinfo  {journal}
  {Reviews of Modern Physics}\ }\textbf {\bibinfo {volume} {92}},\ \bibinfo
  {eid} {030501} (\bibinfo {year} {2020})}\BibitemShut {NoStop}%
\bibitem [{\citenamefont {Brout}\ \emph {et~al.}(2022)\citenamefont {Brout}
  \emph {et~al.}}]{Brout:2022vxf}%
  \BibitemOpen
  \bibfield  {author} {\bibinfo {author} {\bibfnamefont {Dillon}\ \bibnamefont
  {Brout}} \emph {et~al.},\ }\bibfield  {title} {\enquote {\bibinfo {title}
  {{The Pantheon+ Analysis: Cosmological Constraints}},}\ }\href {\doibase
  10.3847/1538-4357/ac8e04} {\bibfield  {journal} {\bibinfo  {journal}
  {Astrophys. J.}\ }\textbf {\bibinfo {volume} {938}},\ \bibinfo {pages} {110}
  (\bibinfo {year} {2022})},\ \Eprint {http://arxiv.org/abs/2202.04077}
  {arXiv:2202.04077 [astro-ph.CO]} \BibitemShut {NoStop}%
\bibitem [{\citenamefont {Aghanim}\ \emph
  {et~al.}(2020{\natexlab{a}})\citenamefont {Aghanim} \emph
  {et~al.}}]{Planck:2018vyg}%
  \BibitemOpen
  \bibfield  {author} {\bibinfo {author} {\bibfnamefont {N.}~\bibnamefont
  {Aghanim}} \emph {et~al.} (\bibinfo {collaboration} {Planck}),\ }\bibfield
  {title} {\enquote {\bibinfo {title} {{Planck 2018 results. VI. Cosmological
  parameters}},}\ }\href {\doibase 10.1051/0004-6361/201833910} {\bibfield
  {journal} {\bibinfo  {journal} {Astron. Astrophys.}\ }\textbf {\bibinfo
  {volume} {641}},\ \bibinfo {pages} {A6} (\bibinfo {year}
  {2020}{\natexlab{a}})},\ \bibinfo {note} {[Erratum: Astron.Astrophys. 652, C4
  (2021)]},\ \Eprint {http://arxiv.org/abs/1807.06209} {arXiv:1807.06209
  [astro-ph.CO]} \BibitemShut {NoStop}%
\bibitem [{\citenamefont {Riess}\ and\ \citenamefont
  {Breuval}(2023)}]{Riess:2023egm}%
  \BibitemOpen
  \bibfield  {author} {\bibinfo {author} {\bibfnamefont {Adam~G.}\ \bibnamefont
  {Riess}}\ and\ \bibinfo {author} {\bibfnamefont {Louise}\ \bibnamefont
  {Breuval}},\ }\bibfield  {title} {\enquote {\bibinfo {title} {{The Local
  Value of H$_0$}},}\ \ }(\bibinfo {year} {2023})\ \Eprint
  {http://arxiv.org/abs/2308.10954} {arXiv:2308.10954 [astro-ph.CO]}
  \BibitemShut {NoStop}%
\bibitem [{\citenamefont {Riess}\ \emph {et~al.}(2023)\citenamefont {Riess},
  \citenamefont {Anand}, \citenamefont {Yuan}, \citenamefont {Casertano},
  \citenamefont {Dolphin}, \citenamefont {Macri}, \citenamefont {Breuval},
  \citenamefont {Scolnic}, \citenamefont {Perrin},\ and\ \citenamefont
  {Anderson}}]{Riess:2023bfx}%
  \BibitemOpen
  \bibfield  {author} {\bibinfo {author} {\bibfnamefont {Adam~G.}\ \bibnamefont
  {Riess}}, \bibinfo {author} {\bibfnamefont {Gagandeep~S.}\ \bibnamefont
  {Anand}}, \bibinfo {author} {\bibfnamefont {Wenlong}\ \bibnamefont {Yuan}},
  \bibinfo {author} {\bibfnamefont {Stefano}\ \bibnamefont {Casertano}},
  \bibinfo {author} {\bibfnamefont {Andrew}\ \bibnamefont {Dolphin}}, \bibinfo
  {author} {\bibfnamefont {Lucas~M.}\ \bibnamefont {Macri}}, \bibinfo {author}
  {\bibfnamefont {Louise}\ \bibnamefont {Breuval}}, \bibinfo {author}
  {\bibfnamefont {Dan}\ \bibnamefont {Scolnic}}, \bibinfo {author}
  {\bibfnamefont {Marshall}\ \bibnamefont {Perrin}}, \ and\ \bibinfo {author}
  {\bibfnamefont {Richard~I.}\ \bibnamefont {Anderson}},\ }\bibfield  {title}
  {\enquote {\bibinfo {title} {{Crowded No More: The Accuracy of the Hubble
  Constant Tested with High Resolution Observations of Cepheids by JWST}},}\
  }\href@noop {} {\  (\bibinfo {year} {2023})},\ \Eprint
  {http://arxiv.org/abs/2307.15806} {arXiv:2307.15806 [astro-ph.CO]}
  \BibitemShut {NoStop}%
\bibitem [{\citenamefont {Sch\"oneberg}\ \emph {et~al.}(2022)\citenamefont
  {Sch\"oneberg}, \citenamefont {Franco~Abell\'an}, \citenamefont
  {P\'erez~S\'anchez}, \citenamefont {Witte}, \citenamefont {Poulin},\ and\
  \citenamefont {Lesgourgues}}]{Schoneberg:2021qvd}%
  \BibitemOpen
  \bibfield  {author} {\bibinfo {author} {\bibfnamefont {Nils}\ \bibnamefont
  {Sch\"oneberg}}, \bibinfo {author} {\bibfnamefont {Guillermo}\ \bibnamefont
  {Franco~Abell\'an}}, \bibinfo {author} {\bibfnamefont {Andrea}\ \bibnamefont
  {P\'erez~S\'anchez}}, \bibinfo {author} {\bibfnamefont {Samuel~J.}\
  \bibnamefont {Witte}}, \bibinfo {author} {\bibfnamefont {Vivian}\
  \bibnamefont {Poulin}}, \ and\ \bibinfo {author} {\bibfnamefont {Julien}\
  \bibnamefont {Lesgourgues}},\ }\bibfield  {title} {\enquote {\bibinfo {title}
  {{The H0 Olympics: A fair ranking of proposed models}},}\ }\href {\doibase
  10.1016/j.physrep.2022.07.001} {\bibfield  {journal} {\bibinfo  {journal}
  {Phys. Rept.}\ }\textbf {\bibinfo {volume} {984}},\ \bibinfo {pages} {1--55}
  (\bibinfo {year} {2022})},\ \Eprint {http://arxiv.org/abs/2107.10291}
  {arXiv:2107.10291 [astro-ph.CO]} \BibitemShut {NoStop}%
\bibitem [{\citenamefont {Di~Valentino}\ \emph {et~al.}(2021)\citenamefont
  {Di~Valentino}, \citenamefont {Mena}, \citenamefont {Pan}, \citenamefont
  {Visinelli}, \citenamefont {Yang}, \citenamefont {Melchiorri}, \citenamefont
  {Mota}, \citenamefont {Riess},\ and\ \citenamefont
  {Silk}}]{DiValentino:2021izs}%
  \BibitemOpen
  \bibfield  {author} {\bibinfo {author} {\bibfnamefont {Eleonora}\
  \bibnamefont {Di~Valentino}}, \bibinfo {author} {\bibfnamefont {Olga}\
  \bibnamefont {Mena}}, \bibinfo {author} {\bibfnamefont {Supriya}\
  \bibnamefont {Pan}}, \bibinfo {author} {\bibfnamefont {Luca}\ \bibnamefont
  {Visinelli}}, \bibinfo {author} {\bibfnamefont {Weiqiang}\ \bibnamefont
  {Yang}}, \bibinfo {author} {\bibfnamefont {Alessandro}\ \bibnamefont
  {Melchiorri}}, \bibinfo {author} {\bibfnamefont {David~F.}\ \bibnamefont
  {Mota}}, \bibinfo {author} {\bibfnamefont {Adam~G.}\ \bibnamefont {Riess}}, \
  and\ \bibinfo {author} {\bibfnamefont {Joseph}\ \bibnamefont {Silk}},\
  }\bibfield  {title} {\enquote {\bibinfo {title} {{In the realm of the Hubble
  tension\textemdash{}a review of solutions}},}\ }\href {\doibase
  10.1088/1361-6382/ac086d} {\bibfield  {journal} {\bibinfo  {journal} {Class.
  Quant. Grav.}\ }\textbf {\bibinfo {volume} {38}},\ \bibinfo {pages} {153001}
  (\bibinfo {year} {2021})},\ \Eprint {http://arxiv.org/abs/2103.01183}
  {arXiv:2103.01183 [astro-ph.CO]} \BibitemShut {NoStop}%
\bibitem [{\citenamefont {Karwal}\ and\ \citenamefont
  {Kamionkowski}(2016)}]{Karwal:2016vyq}%
  \BibitemOpen
  \bibfield  {author} {\bibinfo {author} {\bibfnamefont {Tanvi}\ \bibnamefont
  {Karwal}}\ and\ \bibinfo {author} {\bibfnamefont {Marc}\ \bibnamefont
  {Kamionkowski}},\ }\bibfield  {title} {\enquote {\bibinfo {title} {{Dark
  energy at early times, the Hubble parameter, and the string axiverse}},}\
  }\href {\doibase 10.1103/PhysRevD.94.103523} {\bibfield  {journal} {\bibinfo
  {journal} {Phys. Rev. D}\ }\textbf {\bibinfo {volume} {94}},\ \bibinfo
  {pages} {103523} (\bibinfo {year} {2016})},\ \Eprint
  {http://arxiv.org/abs/1608.01309} {arXiv:1608.01309 [astro-ph.CO]}
  \BibitemShut {NoStop}%
\bibitem [{\citenamefont {Poulin}\ \emph {et~al.}(2019)\citenamefont {Poulin},
  \citenamefont {Smith}, \citenamefont {Karwal},\ and\ \citenamefont
  {Kamionkowski}}]{Poulin:2018cxd}%
  \BibitemOpen
  \bibfield  {author} {\bibinfo {author} {\bibfnamefont {Vivian}\ \bibnamefont
  {Poulin}}, \bibinfo {author} {\bibfnamefont {Tristan~L.}\ \bibnamefont
  {Smith}}, \bibinfo {author} {\bibfnamefont {Tanvi}\ \bibnamefont {Karwal}}, \
  and\ \bibinfo {author} {\bibfnamefont {Marc}\ \bibnamefont {Kamionkowski}},\
  }\bibfield  {title} {\enquote {\bibinfo {title} {{Early Dark Energy Can
  Resolve The Hubble Tension}},}\ }\href {\doibase
  10.1103/PhysRevLett.122.221301} {\bibfield  {journal} {\bibinfo  {journal}
  {Phys. Rev. Lett.}\ }\textbf {\bibinfo {volume} {122}},\ \bibinfo {pages}
  {221301} (\bibinfo {year} {2019})},\ \Eprint
  {http://arxiv.org/abs/1811.04083} {arXiv:1811.04083 [astro-ph.CO]}
  \BibitemShut {NoStop}%
\bibitem [{\citenamefont {Smith}\ \emph {et~al.}(2020)\citenamefont {Smith},
  \citenamefont {Poulin},\ and\ \citenamefont {Amin}}]{Smith:2019ihp}%
  \BibitemOpen
  \bibfield  {author} {\bibinfo {author} {\bibfnamefont {Tristan~L.}\
  \bibnamefont {Smith}}, \bibinfo {author} {\bibfnamefont {Vivian}\
  \bibnamefont {Poulin}}, \ and\ \bibinfo {author} {\bibfnamefont {Mustafa~A.}\
  \bibnamefont {Amin}},\ }\bibfield  {title} {\enquote {\bibinfo {title}
  {{Oscillating scalar fields and the Hubble tension: a resolution with novel
  signatures}},}\ }\href {\doibase 10.1103/PhysRevD.101.063523} {\bibfield
  {journal} {\bibinfo  {journal} {Phys. Rev. D}\ }\textbf {\bibinfo {volume}
  {101}},\ \bibinfo {pages} {063523} (\bibinfo {year} {2020})},\ \Eprint
  {http://arxiv.org/abs/1908.06995} {arXiv:1908.06995 [astro-ph.CO]}
  \BibitemShut {NoStop}%
\bibitem [{\citenamefont {Kamionkowski}\ and\ \citenamefont
  {Riess}(2022)}]{Kamionkowski:2022pkx}%
  \BibitemOpen
  \bibfield  {author} {\bibinfo {author} {\bibfnamefont {Marc}\ \bibnamefont
  {Kamionkowski}}\ and\ \bibinfo {author} {\bibfnamefont {Adam~G.}\
  \bibnamefont {Riess}},\ }\bibfield  {title} {\enquote {\bibinfo {title} {{The
  Hubble Tension and Early Dark Energy}},}\ }\href@noop {} {\  (\bibinfo {year}
  {2022})},\ \Eprint {http://arxiv.org/abs/2211.04492} {arXiv:2211.04492
  [astro-ph.CO]} \BibitemShut {NoStop}%
\bibitem [{\citenamefont {Poulin}\ \emph {et~al.}(2023)\citenamefont {Poulin},
  \citenamefont {Smith},\ and\ \citenamefont {Karwal}}]{Poulin:2023lkg}%
  \BibitemOpen
  \bibfield  {author} {\bibinfo {author} {\bibfnamefont {Vivian}\ \bibnamefont
  {Poulin}}, \bibinfo {author} {\bibfnamefont {Tristan~L.}\ \bibnamefont
  {Smith}}, \ and\ \bibinfo {author} {\bibfnamefont {Tanvi}\ \bibnamefont
  {Karwal}},\ }\bibfield  {title} {\enquote {\bibinfo {title} {{The Ups and
  Downs of Early Dark Energy solutions to the Hubble tension: a review of
  models, hints and constraints circa 2023}},}\ }\href@noop {} {\  (\bibinfo
  {year} {2023})},\ \Eprint {http://arxiv.org/abs/2302.09032} {arXiv:2302.09032
  [astro-ph.CO]} \BibitemShut {NoStop}%
\bibitem [{\citenamefont {Huang}\ \emph {et~al.}(2018)\citenamefont {Huang},
  \citenamefont {Addison}, \citenamefont {Weiland},\ and\ \citenamefont
  {Bennett}}]{Huang:2018xle}%
  \BibitemOpen
  \bibfield  {author} {\bibinfo {author} {\bibfnamefont {Yajing}\ \bibnamefont
  {Huang}}, \bibinfo {author} {\bibfnamefont {Graeme~E.}\ \bibnamefont
  {Addison}}, \bibinfo {author} {\bibfnamefont {Janet~L.}\ \bibnamefont
  {Weiland}}, \ and\ \bibinfo {author} {\bibfnamefont {Charles~L.}\
  \bibnamefont {Bennett}},\ }\bibfield  {title} {\enquote {\bibinfo {title}
  {{Assessing Consistency Between WMAP 9-year and Planck 2015 Temperature Power
  Spectra}},}\ }\href {\doibase 10.3847/1538-4357/aaeb1f} {\bibfield  {journal}
  {\bibinfo  {journal} {Astrophys. J.}\ }\textbf {\bibinfo {volume} {869}},\
  \bibinfo {pages} {38} (\bibinfo {year} {2018})},\ \Eprint
  {http://arxiv.org/abs/1804.05428} {arXiv:1804.05428 [astro-ph.CO]}
  \BibitemShut {NoStop}%
\bibitem [{\citenamefont {Smith}\ \emph {et~al.}(2022)\citenamefont {Smith},
  \citenamefont {Lucca}, \citenamefont {Poulin}, \citenamefont {Abellan},
  \citenamefont {Balkenhol}, \citenamefont {Benabed}, \citenamefont {Galli},\
  and\ \citenamefont {Murgia}}]{Smith:2022hwi}%
  \BibitemOpen
  \bibfield  {author} {\bibinfo {author} {\bibfnamefont {Tristan~L.}\
  \bibnamefont {Smith}}, \bibinfo {author} {\bibfnamefont {Matteo}\
  \bibnamefont {Lucca}}, \bibinfo {author} {\bibfnamefont {Vivian}\
  \bibnamefont {Poulin}}, \bibinfo {author} {\bibfnamefont {Guillermo~F.}\
  \bibnamefont {Abellan}}, \bibinfo {author} {\bibfnamefont {Lennart}\
  \bibnamefont {Balkenhol}}, \bibinfo {author} {\bibfnamefont {Karim}\
  \bibnamefont {Benabed}}, \bibinfo {author} {\bibfnamefont {Silvia}\
  \bibnamefont {Galli}}, \ and\ \bibinfo {author} {\bibfnamefont {Riccardo}\
  \bibnamefont {Murgia}},\ }\bibfield  {title} {\enquote {\bibinfo {title}
  {{Hints of early dark energy in Planck, SPT, and ACT data: New physics or
  systematics?}}}\ }\href {\doibase 10.1103/PhysRevD.106.043526} {\bibfield
  {journal} {\bibinfo  {journal} {Phys. Rev. D}\ }\textbf {\bibinfo {volume}
  {106}},\ \bibinfo {pages} {043526} (\bibinfo {year} {2022})},\ \Eprint
  {http://arxiv.org/abs/2202.09379} {arXiv:2202.09379 [astro-ph.CO]}
  \BibitemShut {NoStop}%
\bibitem [{\citenamefont {Smith}\ \emph {et~al.}(2021)\citenamefont {Smith},
  \citenamefont {Poulin}, \citenamefont {Bernal}, \citenamefont {Boddy},
  \citenamefont {Kamionkowski},\ and\ \citenamefont {Murgia}}]{Smith:2020rxx}%
  \BibitemOpen
  \bibfield  {author} {\bibinfo {author} {\bibfnamefont {Tristan~L.}\
  \bibnamefont {Smith}}, \bibinfo {author} {\bibfnamefont {Vivian}\
  \bibnamefont {Poulin}}, \bibinfo {author} {\bibfnamefont {Jos\'e~Luis}\
  \bibnamefont {Bernal}}, \bibinfo {author} {\bibfnamefont {Kimberly~K.}\
  \bibnamefont {Boddy}}, \bibinfo {author} {\bibfnamefont {Marc}\ \bibnamefont
  {Kamionkowski}}, \ and\ \bibinfo {author} {\bibfnamefont {Riccardo}\
  \bibnamefont {Murgia}},\ }\bibfield  {title} {\enquote {\bibinfo {title}
  {{Early dark energy is not excluded by current large-scale structure
  data}},}\ }\href {\doibase 10.1103/PhysRevD.103.123542} {\bibfield  {journal}
  {\bibinfo  {journal} {Phys. Rev. D}\ }\textbf {\bibinfo {volume} {103}},\
  \bibinfo {pages} {123542} (\bibinfo {year} {2021})},\ \Eprint
  {http://arxiv.org/abs/2009.10740} {arXiv:2009.10740 [astro-ph.CO]}
  \BibitemShut {NoStop}%
\bibitem [{\citenamefont {Herold}\ \emph {et~al.}(2022)\citenamefont {Herold},
  \citenamefont {Ferreira},\ and\ \citenamefont {Komatsu}}]{Herold:2021ksg}%
  \BibitemOpen
  \bibfield  {author} {\bibinfo {author} {\bibfnamefont {Laura}\ \bibnamefont
  {Herold}}, \bibinfo {author} {\bibfnamefont {Elisa G.~M.}\ \bibnamefont
  {Ferreira}}, \ and\ \bibinfo {author} {\bibfnamefont {Eiichiro}\ \bibnamefont
  {Komatsu}},\ }\bibfield  {title} {\enquote {\bibinfo {title} {{New Constraint
  on Early Dark Energy from Planck and BOSS Data Using the Profile
  Likelihood}},}\ }\href {\doibase 10.3847/2041-8213/ac63a3} {\bibfield
  {journal} {\bibinfo  {journal} {Astrophys. J. Lett.}\ }\textbf {\bibinfo
  {volume} {929}},\ \bibinfo {pages} {L16} (\bibinfo {year} {2022})},\ \Eprint
  {http://arxiv.org/abs/2112.12140} {arXiv:2112.12140 [astro-ph.CO]}
  \BibitemShut {NoStop}%
\bibitem [{\citenamefont {Herold}\ and\ \citenamefont
  {Ferreira}(2022)}]{Herold:2022iib}%
  \BibitemOpen
  \bibfield  {author} {\bibinfo {author} {\bibfnamefont {Laura}\ \bibnamefont
  {Herold}}\ and\ \bibinfo {author} {\bibfnamefont {Elisa G.~M.}\ \bibnamefont
  {Ferreira}},\ }\bibfield  {title} {\enquote {\bibinfo {title} {{Resolving the
  Hubble tension with Early Dark Energy}},}\ }\href@noop {} {\  (\bibinfo
  {year} {2022})},\ \Eprint {http://arxiv.org/abs/2210.16296} {arXiv:2210.16296
  [astro-ph.CO]} \BibitemShut {NoStop}%
\bibitem [{\citenamefont {Balkenhol}\ \emph {et~al.}(2022)\citenamefont
  {Balkenhol} \emph {et~al.}}]{SPT-3G:2022hvq}%
  \BibitemOpen
  \bibfield  {author} {\bibinfo {author} {\bibfnamefont {L.}~\bibnamefont
  {Balkenhol}} \emph {et~al.} (\bibinfo {collaboration} {SPT-3G}),\ }\bibfield
  {title} {\enquote {\bibinfo {title} {{A Measurement of the CMB Temperature
  Power Spectrum and Constraints on Cosmology from the SPT-3G 2018 TT/TE/EE
  Data Set}},}\ }\href@noop {} {\  (\bibinfo {year} {2022})},\ \Eprint
  {http://arxiv.org/abs/2212.05642} {arXiv:2212.05642 [astro-ph.CO]}
  \BibitemShut {NoStop}%
\bibitem [{\citenamefont {Franco~Abell\'an}\ \emph {et~al.}(2023)\citenamefont
  {Franco~Abell\'an}, \citenamefont {Braglia}, \citenamefont {Ballardini},
  \citenamefont {Finelli},\ and\ \citenamefont
  {Poulin}}]{FrancoAbellan:2023gec}%
  \BibitemOpen
  \bibfield  {author} {\bibinfo {author} {\bibfnamefont {Guillermo}\
  \bibnamefont {Franco~Abell\'an}}, \bibinfo {author} {\bibfnamefont {Matteo}\
  \bibnamefont {Braglia}}, \bibinfo {author} {\bibfnamefont {Mario}\
  \bibnamefont {Ballardini}}, \bibinfo {author} {\bibfnamefont {Fabio}\
  \bibnamefont {Finelli}}, \ and\ \bibinfo {author} {\bibfnamefont {Vivian}\
  \bibnamefont {Poulin}},\ }\bibfield  {title} {\enquote {\bibinfo {title}
  {{Probing Early Modification of Gravity with Planck, ACT and SPT}},}\
  }\href@noop {} {\  (\bibinfo {year} {2023})},\ \Eprint
  {http://arxiv.org/abs/2308.12345} {arXiv:2308.12345 [astro-ph.CO]}
  \BibitemShut {NoStop}%
\bibitem [{\citenamefont {Hill}\ \emph {et~al.}(2022)\citenamefont {Hill} \emph
  {et~al.}}]{Hill:2021yec}%
  \BibitemOpen
  \bibfield  {author} {\bibinfo {author} {\bibfnamefont {J.~Colin}\
  \bibnamefont {Hill}} \emph {et~al.},\ }\bibfield  {title} {\enquote {\bibinfo
  {title} {{Atacama Cosmology Telescope: Constraints on prerecombination early
  dark energy}},}\ }\href {\doibase 10.1103/PhysRevD.105.123536} {\bibfield
  {journal} {\bibinfo  {journal} {Phys. Rev. D}\ }\textbf {\bibinfo {volume}
  {105}},\ \bibinfo {pages} {123536} (\bibinfo {year} {2022})},\ \Eprint
  {http://arxiv.org/abs/2109.04451} {arXiv:2109.04451 [astro-ph.CO]}
  \BibitemShut {NoStop}%
\bibitem [{\citenamefont {Poulin}\ \emph {et~al.}(2021)\citenamefont {Poulin},
  \citenamefont {Smith},\ and\ \citenamefont {Bartlett}}]{Poulin:2021bjr}%
  \BibitemOpen
  \bibfield  {author} {\bibinfo {author} {\bibfnamefont {Vivian}\ \bibnamefont
  {Poulin}}, \bibinfo {author} {\bibfnamefont {Tristan~L.}\ \bibnamefont
  {Smith}}, \ and\ \bibinfo {author} {\bibfnamefont {Alexa}\ \bibnamefont
  {Bartlett}},\ }\bibfield  {title} {\enquote {\bibinfo {title} {{Dark energy
  at early times and ACT data: A larger Hubble constant without late-time
  priors}},}\ }\href {\doibase 10.1103/PhysRevD.104.123550} {\bibfield
  {journal} {\bibinfo  {journal} {Phys. Rev. D}\ }\textbf {\bibinfo {volume}
  {104}},\ \bibinfo {pages} {123550} (\bibinfo {year} {2021})},\ \Eprint
  {http://arxiv.org/abs/2109.06229} {arXiv:2109.06229 [astro-ph.CO]}
  \BibitemShut {NoStop}%
\bibitem [{\citenamefont {Audren}\ \emph {et~al.}(2013)\citenamefont {Audren},
  \citenamefont {Lesgourgues}, \citenamefont {Benabed},\ and\ \citenamefont
  {Prunet}}]{Audren:2012wb}%
  \BibitemOpen
  \bibfield  {author} {\bibinfo {author} {\bibfnamefont {Benjamin}\
  \bibnamefont {Audren}}, \bibinfo {author} {\bibfnamefont {Julien}\
  \bibnamefont {Lesgourgues}}, \bibinfo {author} {\bibfnamefont {Karim}\
  \bibnamefont {Benabed}}, \ and\ \bibinfo {author} {\bibfnamefont {Simon}\
  \bibnamefont {Prunet}},\ }\bibfield  {title} {\enquote {\bibinfo {title}
  {{Conservative Constraints on Early Cosmology: an illustration of the Monte
  Python cosmological parameter inference code}},}\ }\href {\doibase
  10.1088/1475-7516/2013/02/001} {\bibfield  {journal} {\bibinfo  {journal}
  {JCAP}\ }\textbf {\bibinfo {volume} {02}},\ \bibinfo {pages} {001} (\bibinfo
  {year} {2013})},\ \Eprint {http://arxiv.org/abs/1210.7183} {arXiv:1210.7183
  [astro-ph.CO]} \BibitemShut {NoStop}%
\bibitem [{\citenamefont {Brinckmann}\ and\ \citenamefont
  {Lesgourgues}(2019)}]{Brinckmann:2018cvx}%
  \BibitemOpen
  \bibfield  {author} {\bibinfo {author} {\bibfnamefont {Thejs}\ \bibnamefont
  {Brinckmann}}\ and\ \bibinfo {author} {\bibfnamefont {Julien}\ \bibnamefont
  {Lesgourgues}},\ }\bibfield  {title} {\enquote {\bibinfo {title}
  {{MontePython 3: boosted MCMC sampler and other features}},}\ }\href
  {\doibase 10.1016/j.dark.2018.100260} {\bibfield  {journal} {\bibinfo
  {journal} {Phys. Dark Univ.}\ }\textbf {\bibinfo {volume} {24}},\ \bibinfo
  {pages} {100260} (\bibinfo {year} {2019})},\ \Eprint
  {http://arxiv.org/abs/1804.07261} {arXiv:1804.07261 [astro-ph.CO]}
  \BibitemShut {NoStop}%
\bibitem [{\citenamefont {Lesgourgues}(2011)}]{Lesgourgues:2011re}%
  \BibitemOpen
  \bibfield  {author} {\bibinfo {author} {\bibfnamefont {Julien}\ \bibnamefont
  {Lesgourgues}},\ }\bibfield  {title} {\enquote {\bibinfo {title} {{The Cosmic
  Linear Anisotropy Solving System (CLASS) I: Overview}},}\ }\href@noop {} {\
  (\bibinfo {year} {2011})},\ \Eprint {http://arxiv.org/abs/1104.2932}
  {arXiv:1104.2932 [astro-ph.IM]} \BibitemShut {NoStop}%
\bibitem [{\citenamefont {Blas}\ \emph {et~al.}(2011)\citenamefont {Blas},
  \citenamefont {Lesgourgues},\ and\ \citenamefont {Tram}}]{Blas:2011rf}%
  \BibitemOpen
  \bibfield  {author} {\bibinfo {author} {\bibfnamefont {Diego}\ \bibnamefont
  {Blas}}, \bibinfo {author} {\bibfnamefont {Julien}\ \bibnamefont
  {Lesgourgues}}, \ and\ \bibinfo {author} {\bibfnamefont {Thomas}\
  \bibnamefont {Tram}},\ }\bibfield  {title} {\enquote {\bibinfo {title} {{The
  Cosmic Linear Anisotropy Solving System (CLASS) II: Approximation
  schemes}},}\ }\href {\doibase 10.1088/1475-7516/2011/07/034} {\bibfield
  {journal} {\bibinfo  {journal} {JCAP}\ }\textbf {\bibinfo {volume} {07}},\
  \bibinfo {pages} {034} (\bibinfo {year} {2011})},\ \Eprint
  {http://arxiv.org/abs/1104.2933} {arXiv:1104.2933 [astro-ph.CO]} \BibitemShut
  {NoStop}%
\bibitem [{\citenamefont {Smith}\ \emph {et~al.}(2003)\citenamefont {Smith},
  \citenamefont {Peacock}, \citenamefont {Jenkins}, \citenamefont {White},
  \citenamefont {Frenk}, \citenamefont {Pearce}, \citenamefont {Thomas},
  \citenamefont {Efstathiou},\ and\ \citenamefont {Couchmann}}]{Smith:2002dz}%
  \BibitemOpen
  \bibfield  {author} {\bibinfo {author} {\bibfnamefont {R.~E.}\ \bibnamefont
  {Smith}}, \bibinfo {author} {\bibfnamefont {J.~A.}\ \bibnamefont {Peacock}},
  \bibinfo {author} {\bibfnamefont {A.}~\bibnamefont {Jenkins}}, \bibinfo
  {author} {\bibfnamefont {S.~D.~M.}\ \bibnamefont {White}}, \bibinfo {author}
  {\bibfnamefont {C.~S.}\ \bibnamefont {Frenk}}, \bibinfo {author}
  {\bibfnamefont {F.~R.}\ \bibnamefont {Pearce}}, \bibinfo {author}
  {\bibfnamefont {P.~A.}\ \bibnamefont {Thomas}}, \bibinfo {author}
  {\bibfnamefont {G.}~\bibnamefont {Efstathiou}}, \ and\ \bibinfo {author}
  {\bibfnamefont {H.~M.~P.}\ \bibnamefont {Couchmann}} (\bibinfo
  {collaboration} {VIRGO Consortium}),\ }\bibfield  {title} {\enquote {\bibinfo
  {title} {{Stable clustering, the halo model and nonlinear cosmological power
  spectra}},}\ }\href {\doibase 10.1046/j.1365-8711.2003.06503.x} {\bibfield
  {journal} {\bibinfo  {journal} {Mon. Not. Roy. Astron. Soc.}\ }\textbf
  {\bibinfo {volume} {341}},\ \bibinfo {pages} {1311} (\bibinfo {year}
  {2003})},\ \Eprint {http://arxiv.org/abs/astro-ph/0207664}
  {arXiv:astro-ph/0207664} \BibitemShut {NoStop}%
\bibitem [{\citenamefont {Gelman}\ and\ \citenamefont
  {Rubin}(1992)}]{Gelman:1992zz}%
  \BibitemOpen
  \bibfield  {author} {\bibinfo {author} {\bibfnamefont {Andrew}\ \bibnamefont
  {Gelman}}\ and\ \bibinfo {author} {\bibfnamefont {Donald~B.}\ \bibnamefont
  {Rubin}},\ }\bibfield  {title} {\enquote {\bibinfo {title} {{Inference from
  Iterative Simulation Using Multiple Sequences}},}\ }\href {\doibase
  10.1214/ss/1177011136} {\bibfield  {journal} {\bibinfo  {journal} {Statist.
  Sci.}\ }\textbf {\bibinfo {volume} {7}},\ \bibinfo {pages} {457--472}
  (\bibinfo {year} {1992})}\BibitemShut {NoStop}%
\bibitem [{\citenamefont {Lewis}(2019)}]{Lewis:2019xzd}%
  \BibitemOpen
  \bibfield  {author} {\bibinfo {author} {\bibfnamefont {Antony}\ \bibnamefont
  {Lewis}},\ }\bibfield  {title} {\enquote {\bibinfo {title} {{GetDist: a
  Python package for analysing Monte Carlo samples}},}\ }\href@noop {} {\
  (\bibinfo {year} {2019})},\ \Eprint {http://arxiv.org/abs/1910.13970}
  {arXiv:1910.13970 [astro-ph.IM]} \BibitemShut {NoStop}%
\bibitem [{\citenamefont {Aghanim}\ \emph
  {et~al.}(2020{\natexlab{b}})\citenamefont {Aghanim} \emph
  {et~al.}}]{Planck:2019nip}%
  \BibitemOpen
  \bibfield  {author} {\bibinfo {author} {\bibfnamefont {N.}~\bibnamefont
  {Aghanim}} \emph {et~al.} (\bibinfo {collaboration} {Planck}),\ }\bibfield
  {title} {\enquote {\bibinfo {title} {{Planck 2018 results. V. CMB power
  spectra and likelihoods}},}\ }\href {\doibase 10.1051/0004-6361/201936386}
  {\bibfield  {journal} {\bibinfo  {journal} {Astron. Astrophys.}\ }\textbf
  {\bibinfo {volume} {641}},\ \bibinfo {pages} {A5} (\bibinfo {year}
  {2020}{\natexlab{b}})},\ \Eprint {http://arxiv.org/abs/1907.12875}
  {arXiv:1907.12875 [astro-ph.CO]} \BibitemShut {NoStop}%
\bibitem [{\citenamefont {Aghanim}\ \emph
  {et~al.}(2020{\natexlab{c}})\citenamefont {Aghanim} \emph
  {et~al.}}]{Planck:2018lbu}%
  \BibitemOpen
  \bibfield  {author} {\bibinfo {author} {\bibfnamefont {N.}~\bibnamefont
  {Aghanim}} \emph {et~al.} (\bibinfo {collaboration} {Planck}),\ }\bibfield
  {title} {\enquote {\bibinfo {title} {{Planck 2018 results. VIII.
  Gravitational lensing}},}\ }\href {\doibase 10.1051/0004-6361/201833886}
  {\bibfield  {journal} {\bibinfo  {journal} {Astron. Astrophys.}\ }\textbf
  {\bibinfo {volume} {641}},\ \bibinfo {pages} {A8} (\bibinfo {year}
  {2020}{\natexlab{c}})},\ \Eprint {http://arxiv.org/abs/1807.06210}
  {arXiv:1807.06210 [astro-ph.CO]} \BibitemShut {NoStop}%
\bibitem [{\citenamefont {Dutcher}\ \emph {et~al.}(2021)\citenamefont {Dutcher}
  \emph {et~al.}}]{SPT-3G:2021eoc}%
  \BibitemOpen
  \bibfield  {author} {\bibinfo {author} {\bibfnamefont {D.}~\bibnamefont
  {Dutcher}} \emph {et~al.} (\bibinfo {collaboration} {SPT-3G}),\ }\bibfield
  {title} {\enquote {\bibinfo {title} {{Measurements of the E-mode polarization
  and temperature-E-mode correlation of the CMB from SPT-3G 2018 data}},}\
  }\href {\doibase 10.1103/PhysRevD.104.022003} {\bibfield  {journal} {\bibinfo
   {journal} {Phys. Rev. D}\ }\textbf {\bibinfo {volume} {104}},\ \bibinfo
  {pages} {022003} (\bibinfo {year} {2021})},\ \Eprint
  {http://arxiv.org/abs/2101.01684} {arXiv:2101.01684 [astro-ph.CO]}
  \BibitemShut {NoStop}%
\bibitem [{\citenamefont {Choi}\ \emph {et~al.}(2020)\citenamefont {Choi} \emph
  {et~al.}}]{ACT:2020frw}%
  \BibitemOpen
  \bibfield  {author} {\bibinfo {author} {\bibfnamefont {Steve~K.}\
  \bibnamefont {Choi}} \emph {et~al.} (\bibinfo {collaboration} {ACT}),\
  }\bibfield  {title} {\enquote {\bibinfo {title} {{The Atacama Cosmology
  Telescope: a measurement of the Cosmic Microwave Background power spectra at
  98 and 150 GHz}},}\ }\href {\doibase 10.1088/1475-7516/2020/12/045}
  {\bibfield  {journal} {\bibinfo  {journal} {JCAP}\ }\textbf {\bibinfo
  {volume} {12}},\ \bibinfo {pages} {045} (\bibinfo {year} {2020})},\ \Eprint
  {http://arxiv.org/abs/2007.07289} {arXiv:2007.07289 [astro-ph.CO]}
  \BibitemShut {NoStop}%
\bibitem [{\citenamefont {Aiola}\ \emph {et~al.}(2020)\citenamefont {Aiola}
  \emph {et~al.}}]{ACT:2020gnv}%
  \BibitemOpen
  \bibfield  {author} {\bibinfo {author} {\bibfnamefont {Simone}\ \bibnamefont
  {Aiola}} \emph {et~al.} (\bibinfo {collaboration} {ACT}),\ }\bibfield
  {title} {\enquote {\bibinfo {title} {{The Atacama Cosmology Telescope: DR4
  Maps and Cosmological Parameters}},}\ }\href {\doibase
  10.1088/1475-7516/2020/12/047} {\bibfield  {journal} {\bibinfo  {journal}
  {JCAP}\ }\textbf {\bibinfo {volume} {12}},\ \bibinfo {pages} {047} (\bibinfo
  {year} {2020})},\ \Eprint {http://arxiv.org/abs/2007.07288} {arXiv:2007.07288
  [astro-ph.CO]} \BibitemShut {NoStop}%
\bibitem [{\citenamefont {Ross}\ \emph {et~al.}(2015)\citenamefont {Ross},
  \citenamefont {Samushia}, \citenamefont {Howlett}, \citenamefont {Percival},
  \citenamefont {Burden},\ and\ \citenamefont {Manera}}]{Ross:2014qpa}%
  \BibitemOpen
  \bibfield  {author} {\bibinfo {author} {\bibfnamefont {Ashley~J.}\
  \bibnamefont {Ross}}, \bibinfo {author} {\bibfnamefont {Lado}\ \bibnamefont
  {Samushia}}, \bibinfo {author} {\bibfnamefont {Cullan}\ \bibnamefont
  {Howlett}}, \bibinfo {author} {\bibfnamefont {Will~J.}\ \bibnamefont
  {Percival}}, \bibinfo {author} {\bibfnamefont {Angela}\ \bibnamefont
  {Burden}}, \ and\ \bibinfo {author} {\bibfnamefont {Marc}\ \bibnamefont
  {Manera}},\ }\bibfield  {title} {\enquote {\bibinfo {title} {{The clustering
  of the SDSS DR7 main Galaxy sample \textendash{} I. A 4 per cent distance
  measure at $z = 0.15$}},}\ }\href {\doibase 10.1093/mnras/stv154} {\bibfield
  {journal} {\bibinfo  {journal} {Mon. Not. Roy. Astron. Soc.}\ }\textbf
  {\bibinfo {volume} {449}},\ \bibinfo {pages} {835--847} (\bibinfo {year}
  {2015})},\ \Eprint {http://arxiv.org/abs/1409.3242} {arXiv:1409.3242
  [astro-ph.CO]} \BibitemShut {NoStop}%
\bibitem [{\citenamefont {Alam}\ \emph {et~al.}(2017)\citenamefont {Alam} \emph
  {et~al.}}]{BOSS:2016wmc}%
  \BibitemOpen
  \bibfield  {author} {\bibinfo {author} {\bibfnamefont {Shadab}\ \bibnamefont
  {Alam}} \emph {et~al.} (\bibinfo {collaboration} {BOSS}),\ }\bibfield
  {title} {\enquote {\bibinfo {title} {{The clustering of galaxies in the
  completed SDSS-III Baryon Oscillation Spectroscopic Survey: cosmological
  analysis of the DR12 galaxy sample}},}\ }\href {\doibase
  10.1093/mnras/stx721} {\bibfield  {journal} {\bibinfo  {journal} {Mon. Not.
  Roy. Astron. Soc.}\ }\textbf {\bibinfo {volume} {470}},\ \bibinfo {pages}
  {2617--2652} (\bibinfo {year} {2017})},\ \Eprint
  {http://arxiv.org/abs/1607.03155} {arXiv:1607.03155 [astro-ph.CO]}
  \BibitemShut {NoStop}%
\bibitem [{\citenamefont {Riess}\ \emph {et~al.}(2022)\citenamefont {Riess}
  \emph {et~al.}}]{Riess:2021jrx}%
  \BibitemOpen
  \bibfield  {author} {\bibinfo {author} {\bibfnamefont {Adam~G.}\ \bibnamefont
  {Riess}} \emph {et~al.},\ }\bibfield  {title} {\enquote {\bibinfo {title} {{A
  Comprehensive Measurement of the Local Value of the Hubble Constant with 1
  km/s/Mpcå Uncertainty from the Hubble Space Telescope and the SH0ES
  Team}},}\ }\href {\doibase 10.3847/2041-8213/ac5c5b} {\bibfield  {journal}
  {\bibinfo  {journal} {Astrophys. J. Lett.}\ }\textbf {\bibinfo {volume}
  {934}},\ \bibinfo {pages} {L7} (\bibinfo {year} {2022})},\ \Eprint
  {http://arxiv.org/abs/2112.04510} {arXiv:2112.04510 [astro-ph.CO]}
  \BibitemShut {NoStop}%
\bibitem [{\citenamefont {Hu}\ \emph {et~al.}(2001)\citenamefont {Hu},
  \citenamefont {Fukugita}, \citenamefont {Zaldarriaga},\ and\ \citenamefont
  {Tegmark}}]{Hu:2000ti}%
  \BibitemOpen
  \bibfield  {author} {\bibinfo {author} {\bibfnamefont {Wayne}\ \bibnamefont
  {Hu}}, \bibinfo {author} {\bibfnamefont {Masataka}\ \bibnamefont {Fukugita}},
  \bibinfo {author} {\bibfnamefont {Matias}\ \bibnamefont {Zaldarriaga}}, \
  and\ \bibinfo {author} {\bibfnamefont {Max}\ \bibnamefont {Tegmark}},\
  }\bibfield  {title} {\enquote {\bibinfo {title} {{CMB observables and their
  cosmological implications}},}\ }\href {\doibase 10.1086/319449} {\bibfield
  {journal} {\bibinfo  {journal} {Astrophys. J.}\ }\textbf {\bibinfo {volume}
  {549}},\ \bibinfo {pages} {669} (\bibinfo {year} {2001})},\ \Eprint
  {http://arxiv.org/abs/astro-ph/0006436} {arXiv:astro-ph/0006436} \BibitemShut
  {NoStop}%
\bibitem [{\citenamefont {Silk}(1968)}]{Silk:1967kq}%
  \BibitemOpen
  \bibfield  {author} {\bibinfo {author} {\bibfnamefont {Joseph}\ \bibnamefont
  {Silk}},\ }\bibfield  {title} {\enquote {\bibinfo {title} {{Cosmic black body
  radiation and galaxy formation}},}\ }\href {\doibase 10.1086/149449}
  {\bibfield  {journal} {\bibinfo  {journal} {Astrophys. J.}\ }\textbf
  {\bibinfo {volume} {151}},\ \bibinfo {pages} {459--471} (\bibinfo {year}
  {1968})}\BibitemShut {NoStop}%
\bibitem [{\citenamefont {Ade}\ \emph {et~al.}(2014)\citenamefont {Ade} \emph
  {et~al.}}]{Planck:2013pxb}%
  \BibitemOpen
  \bibfield  {author} {\bibinfo {author} {\bibfnamefont {P.~A.~R.}\
  \bibnamefont {Ade}} \emph {et~al.} (\bibinfo {collaboration} {Planck}),\
  }\bibfield  {title} {\enquote {\bibinfo {title} {{Planck 2013 results. XVI.
  Cosmological parameters}},}\ }\href {\doibase 10.1051/0004-6361/201321591}
  {\bibfield  {journal} {\bibinfo  {journal} {Astron. Astrophys.}\ }\textbf
  {\bibinfo {volume} {571}},\ \bibinfo {pages} {A16} (\bibinfo {year}
  {2014})},\ \Eprint {http://arxiv.org/abs/1303.5076} {arXiv:1303.5076
  [astro-ph.CO]} \BibitemShut {NoStop}%
\bibitem [{\citenamefont {Addison}(2021)}]{Addison:2021amj}%
  \BibitemOpen
  \bibfield  {author} {\bibinfo {author} {\bibfnamefont {Graeme~E.}\
  \bibnamefont {Addison}},\ }\bibfield  {title} {\enquote {\bibinfo {title}
  {{High $H_0$ Values from CMB E-mode Data: A Clue for Resolving the Hubble
  Tension?}}}\ }\href {\doibase 10.3847/2041-8213/abf56e} {\bibfield  {journal}
  {\bibinfo  {journal} {Astrophys. J. Lett.}\ }\textbf {\bibinfo {volume}
  {912}},\ \bibinfo {pages} {L1} (\bibinfo {year} {2021})},\ \Eprint
  {http://arxiv.org/abs/2102.00028} {arXiv:2102.00028 [astro-ph.CO]}
  \BibitemShut {NoStop}%
\bibitem [{\citenamefont {Handley}\ and\ \citenamefont
  {Lemos}(2021)}]{Handley:2020hdp}%
  \BibitemOpen
  \bibfield  {author} {\bibinfo {author} {\bibfnamefont {Will}\ \bibnamefont
  {Handley}}\ and\ \bibinfo {author} {\bibfnamefont {Pablo}\ \bibnamefont
  {Lemos}},\ }\bibfield  {title} {\enquote {\bibinfo {title} {{Quantifying the
  global parameter tensions between ACT, SPT and Planck}},}\ }\href {\doibase
  10.1103/PhysRevD.103.063529} {\bibfield  {journal} {\bibinfo  {journal}
  {Phys. Rev. D}\ }\textbf {\bibinfo {volume} {103}},\ \bibinfo {pages}
  {063529} (\bibinfo {year} {2021})},\ \Eprint
  {http://arxiv.org/abs/2007.08496} {arXiv:2007.08496 [astro-ph.CO]}
  \BibitemShut {NoStop}%
\bibitem [{\citenamefont {Hill}\ \emph {et~al.}(2020)\citenamefont {Hill},
  \citenamefont {McDonough}, \citenamefont {Toomey},\ and\ \citenamefont
  {Alexander}}]{Hill:2020osr}%
  \BibitemOpen
  \bibfield  {author} {\bibinfo {author} {\bibfnamefont {J.~Colin}\
  \bibnamefont {Hill}}, \bibinfo {author} {\bibfnamefont {Evan}\ \bibnamefont
  {McDonough}}, \bibinfo {author} {\bibfnamefont {Michael~W.}\ \bibnamefont
  {Toomey}}, \ and\ \bibinfo {author} {\bibfnamefont {Stephon}\ \bibnamefont
  {Alexander}},\ }\bibfield  {title} {\enquote {\bibinfo {title} {{Early dark
  energy does not restore cosmological concordance}},}\ }\href {\doibase
  10.1103/PhysRevD.102.043507} {\bibfield  {journal} {\bibinfo  {journal}
  {Phys. Rev. D}\ }\textbf {\bibinfo {volume} {102}},\ \bibinfo {pages}
  {043507} (\bibinfo {year} {2020})},\ \Eprint
  {http://arxiv.org/abs/2003.07355} {arXiv:2003.07355 [astro-ph.CO]}
  \BibitemShut {NoStop}%
\bibitem [{\citenamefont {Simon}\ \emph {et~al.}(2023)\citenamefont {Simon},
  \citenamefont {Zhang}, \citenamefont {Poulin},\ and\ \citenamefont
  {Smith}}]{Simon:2022adh}%
  \BibitemOpen
  \bibfield  {author} {\bibinfo {author} {\bibfnamefont {Th\'eo}\ \bibnamefont
  {Simon}}, \bibinfo {author} {\bibfnamefont {Pierre}\ \bibnamefont {Zhang}},
  \bibinfo {author} {\bibfnamefont {Vivian}\ \bibnamefont {Poulin}}, \ and\
  \bibinfo {author} {\bibfnamefont {Tristan~L.}\ \bibnamefont {Smith}},\
  }\bibfield  {title} {\enquote {\bibinfo {title} {{Updated constraints from
  the effective field theory analysis of the BOSS power spectrum on early dark
  energy}},}\ }\href {\doibase 10.1103/PhysRevD.107.063505} {\bibfield
  {journal} {\bibinfo  {journal} {Phys. Rev. D}\ }\textbf {\bibinfo {volume}
  {107}},\ \bibinfo {pages} {063505} (\bibinfo {year} {2023})},\ \Eprint
  {http://arxiv.org/abs/2208.05930} {arXiv:2208.05930 [astro-ph.CO]}
  \BibitemShut {NoStop}%
\bibitem [{\citenamefont {Raveri}\ and\ \citenamefont
  {Doux}(2021)}]{Raveri:2021wfz}%
  \BibitemOpen
  \bibfield  {author} {\bibinfo {author} {\bibfnamefont {Marco}\ \bibnamefont
  {Raveri}}\ and\ \bibinfo {author} {\bibfnamefont {Cyrille}\ \bibnamefont
  {Doux}},\ }\bibfield  {title} {\enquote {\bibinfo {title} {{Non-Gaussian
  estimates of tensions in cosmological parameters}},}\ }\href {\doibase
  10.1103/PhysRevD.104.043504} {\bibfield  {journal} {\bibinfo  {journal}
  {Phys. Rev. D}\ }\textbf {\bibinfo {volume} {104}},\ \bibinfo {pages}
  {043504} (\bibinfo {year} {2021})},\ \Eprint
  {http://arxiv.org/abs/2105.03324} {arXiv:2105.03324 [astro-ph.CO]}
  \BibitemShut {NoStop}%
\bibitem [{\citenamefont {Eisenstein}\ \emph {et~al.}(1999)\citenamefont
  {Eisenstein}, \citenamefont {Hu},\ and\ \citenamefont
  {Tegmark}}]{Eisenstein:1998hr}%
  \BibitemOpen
  \bibfield  {author} {\bibinfo {author} {\bibfnamefont {Daniel~J.}\
  \bibnamefont {Eisenstein}}, \bibinfo {author} {\bibfnamefont {Wayne}\
  \bibnamefont {Hu}}, \ and\ \bibinfo {author} {\bibfnamefont {Max}\
  \bibnamefont {Tegmark}},\ }\bibfield  {title} {\enquote {\bibinfo {title}
  {{Cosmic complementarity: Joint parameter estimation from CMB experiments and
  redshift surveys}},}\ }\href {\doibase 10.1086/307261} {\bibfield  {journal}
  {\bibinfo  {journal} {Astrophys. J.}\ }\textbf {\bibinfo {volume} {518}},\
  \bibinfo {pages} {2--23} (\bibinfo {year} {1999})},\ \Eprint
  {http://arxiv.org/abs/astro-ph/9807130} {arXiv:astro-ph/9807130} \BibitemShut
  {NoStop}%
\bibitem [{\citenamefont {Colombo}\ \emph {et~al.}(2009)\citenamefont
  {Colombo}, \citenamefont {Pierpaoli},\ and\ \citenamefont
  {Pritchard}}]{Colombo:2008ta}%
  \BibitemOpen
  \bibfield  {author} {\bibinfo {author} {\bibfnamefont {L.~P.~L.}\
  \bibnamefont {Colombo}}, \bibinfo {author} {\bibfnamefont {E.}~\bibnamefont
  {Pierpaoli}}, \ and\ \bibinfo {author} {\bibfnamefont {J.~R.}\ \bibnamefont
  {Pritchard}},\ }\bibfield  {title} {\enquote {\bibinfo {title} {{Cosmological
  parameters after WMAP5: forecasts for Planck and future galaxy surveys}},}\
  }\href {\doibase 10.1111/j.1365-2966.2009.14802.x} {\bibfield  {journal}
  {\bibinfo  {journal} {Mon. Not. Roy. Astron. Soc.}\ }\textbf {\bibinfo
  {volume} {398}},\ \bibinfo {pages} {1621} (\bibinfo {year} {2009})},\ \Eprint
  {http://arxiv.org/abs/0811.2622} {arXiv:0811.2622 [astro-ph]} \BibitemShut
  {NoStop}%
\bibitem [{\citenamefont {Galli}\ \emph {et~al.}(2010)\citenamefont {Galli},
  \citenamefont {Martinelli}, \citenamefont {Melchiorri}, \citenamefont
  {Pagano}, \citenamefont {Sherwin},\ and\ \citenamefont
  {Spergel}}]{Galli:2010it}%
  \BibitemOpen
  \bibfield  {author} {\bibinfo {author} {\bibfnamefont {Silvia}\ \bibnamefont
  {Galli}}, \bibinfo {author} {\bibfnamefont {Matteo}\ \bibnamefont
  {Martinelli}}, \bibinfo {author} {\bibfnamefont {Alessandro}\ \bibnamefont
  {Melchiorri}}, \bibinfo {author} {\bibfnamefont {Luca}\ \bibnamefont
  {Pagano}}, \bibinfo {author} {\bibfnamefont {Blake~D.}\ \bibnamefont
  {Sherwin}}, \ and\ \bibinfo {author} {\bibfnamefont {David~N.}\ \bibnamefont
  {Spergel}},\ }\bibfield  {title} {\enquote {\bibinfo {title} {{Constraining
  Fundamental Physics with Future CMB Experiments}},}\ }\href {\doibase
  10.1103/PhysRevD.82.123504} {\bibfield  {journal} {\bibinfo  {journal} {Phys.
  Rev. D}\ }\textbf {\bibinfo {volume} {82}},\ \bibinfo {pages} {123504}
  (\bibinfo {year} {2010})},\ \Eprint {http://arxiv.org/abs/1005.3808}
  {arXiv:1005.3808 [astro-ph.CO]} \BibitemShut {NoStop}%
\bibitem [{\citenamefont {Wu}\ \emph {et~al.}(2014)\citenamefont {Wu},
  \citenamefont {Errard}, \citenamefont {Dvorkin}, \citenamefont {Kuo},
  \citenamefont {Lee}, \citenamefont {McDonald}, \citenamefont {Slosar},\ and\
  \citenamefont {Zahn}}]{Wu:2014hta}%
  \BibitemOpen
  \bibfield  {author} {\bibinfo {author} {\bibfnamefont {W.~L.~K.}\
  \bibnamefont {Wu}}, \bibinfo {author} {\bibfnamefont {J.}~\bibnamefont
  {Errard}}, \bibinfo {author} {\bibfnamefont {C.}~\bibnamefont {Dvorkin}},
  \bibinfo {author} {\bibfnamefont {C.~L.}\ \bibnamefont {Kuo}}, \bibinfo
  {author} {\bibfnamefont {A.~T.}\ \bibnamefont {Lee}}, \bibinfo {author}
  {\bibfnamefont {P.}~\bibnamefont {McDonald}}, \bibinfo {author}
  {\bibfnamefont {A.}~\bibnamefont {Slosar}}, \ and\ \bibinfo {author}
  {\bibfnamefont {O.}~\bibnamefont {Zahn}},\ }\bibfield  {title} {\enquote
  {\bibinfo {title} {{A Guide to Designing Future Ground-based Cosmic Microwave
  Background Experiments}},}\ }\href {\doibase 10.1088/0004-637X/788/2/138}
  {\bibfield  {journal} {\bibinfo  {journal} {Astrophys. J.}\ }\textbf
  {\bibinfo {volume} {788}},\ \bibinfo {pages} {138} (\bibinfo {year}
  {2014})},\ \Eprint {http://arxiv.org/abs/1402.4108} {arXiv:1402.4108
  [astro-ph.CO]} \BibitemShut {NoStop}%
\bibitem [{\citenamefont {Rocha}\ \emph {et~al.}(2004)\citenamefont {Rocha},
  \citenamefont {Trotta}, \citenamefont {Martins}, \citenamefont {Melchiorri},
  \citenamefont {Avelino}, \citenamefont {Bean},\ and\ \citenamefont
  {Viana}}]{Rocha:2003gc}%
  \BibitemOpen
  \bibfield  {author} {\bibinfo {author} {\bibfnamefont {G.}~\bibnamefont
  {Rocha}}, \bibinfo {author} {\bibfnamefont {R.}~\bibnamefont {Trotta}},
  \bibinfo {author} {\bibfnamefont {C.~J. A.~P.}\ \bibnamefont {Martins}},
  \bibinfo {author} {\bibfnamefont {A.}~\bibnamefont {Melchiorri}}, \bibinfo
  {author} {\bibfnamefont {P.~P.}\ \bibnamefont {Avelino}}, \bibinfo {author}
  {\bibfnamefont {R.}~\bibnamefont {Bean}}, \ and\ \bibinfo {author}
  {\bibfnamefont {Pedro T.~P.}\ \bibnamefont {Viana}},\ }\bibfield  {title}
  {\enquote {\bibinfo {title} {{Measuring alpha in the early universe: cmb
  polarization, reionization and the fisher matrix analysis}},}\ }\href
  {\doibase 10.1111/j.1365-2966.2004.07832.x} {\bibfield  {journal} {\bibinfo
  {journal} {Mon. Not. Roy. Astron. Soc.}\ }\textbf {\bibinfo {volume} {352}},\
  \bibinfo {pages} {20} (\bibinfo {year} {2004})},\ \Eprint
  {http://arxiv.org/abs/astro-ph/0309211} {arXiv:astro-ph/0309211} \BibitemShut
  {NoStop}%
\bibitem [{\citenamefont {Galli}\ \emph {et~al.}(2014)\citenamefont {Galli},
  \citenamefont {Benabed}, \citenamefont {Bouchet}, \citenamefont {Cardoso},
  \citenamefont {Elsner}, \citenamefont {Hivon}, \citenamefont {Mangilli},
  \citenamefont {Prunet},\ and\ \citenamefont {Wandelt}}]{Galli:2014kla}%
  \BibitemOpen
  \bibfield  {author} {\bibinfo {author} {\bibfnamefont {Silvia}\ \bibnamefont
  {Galli}}, \bibinfo {author} {\bibfnamefont {Karim}\ \bibnamefont {Benabed}},
  \bibinfo {author} {\bibfnamefont {Fran\c{c}ois}\ \bibnamefont {Bouchet}},
  \bibinfo {author} {\bibfnamefont {Jean-Fran\c{c}ois}\ \bibnamefont
  {Cardoso}}, \bibinfo {author} {\bibfnamefont {Franz}\ \bibnamefont {Elsner}},
  \bibinfo {author} {\bibfnamefont {Eric}\ \bibnamefont {Hivon}}, \bibinfo
  {author} {\bibfnamefont {Anna}\ \bibnamefont {Mangilli}}, \bibinfo {author}
  {\bibfnamefont {Simon}\ \bibnamefont {Prunet}}, \ and\ \bibinfo {author}
  {\bibfnamefont {Benjamin}\ \bibnamefont {Wandelt}},\ }\bibfield  {title}
  {\enquote {\bibinfo {title} {{CMB Polarization can constrain cosmology better
  than CMB temperature}},}\ }\href {\doibase 10.1103/PhysRevD.90.063504}
  {\bibfield  {journal} {\bibinfo  {journal} {Phys. Rev. D}\ }\textbf {\bibinfo
  {volume} {90}},\ \bibinfo {pages} {063504} (\bibinfo {year} {2014})},\
  \Eprint {http://arxiv.org/abs/1403.5271} {arXiv:1403.5271 [astro-ph.CO]}
  \BibitemShut {NoStop}%
\bibitem [{\citenamefont {Tucci}\ and\ \citenamefont
  {Toffolatti}(2012)}]{Tucci_2012}%
  \BibitemOpen
  \bibfield  {author} {\bibinfo {author} {\bibfnamefont {Marco}\ \bibnamefont
  {Tucci}}\ and\ \bibinfo {author} {\bibfnamefont {Luigi}\ \bibnamefont
  {Toffolatti}},\ }\bibfield  {title} {\enquote {\bibinfo {title} {The impact
  of polarized extragalactic radio sources on the detection of {CMB}
  anisotropies in polarization},}\ }\href {\doibase 10.1155/2012/624987}
  {\bibfield  {journal} {\bibinfo  {journal} {Advances in Astronomy}\ }\textbf
  {\bibinfo {volume} {2012}},\ \bibinfo {pages} {1--17} (\bibinfo {year}
  {2012})}\BibitemShut {NoStop}%
\bibitem [{\citenamefont {Raveri}\ and\ \citenamefont
  {Hu}(2019)}]{Raveri:2018wln}%
  \BibitemOpen
  \bibfield  {author} {\bibinfo {author} {\bibfnamefont {Marco}\ \bibnamefont
  {Raveri}}\ and\ \bibinfo {author} {\bibfnamefont {Wayne}\ \bibnamefont
  {Hu}},\ }\bibfield  {title} {\enquote {\bibinfo {title} {{Concordance and
  Discordance in Cosmology}},}\ }\href {\doibase 10.1103/PhysRevD.99.043506}
  {\bibfield  {journal} {\bibinfo  {journal} {Phys. Rev. D}\ }\textbf {\bibinfo
  {volume} {99}},\ \bibinfo {pages} {043506} (\bibinfo {year} {2019})},\
  \Eprint {http://arxiv.org/abs/1806.04649} {arXiv:1806.04649 [astro-ph.CO]}
  \BibitemShut {NoStop}%
\bibitem [{\citenamefont {Schöneberg}\ \emph {et~al.}(2019)\citenamefont
  {Schöneberg}, \citenamefont {Lesgourgues},\ and\ \citenamefont
  {Hooper}}]{Schoneberg_2019}%
  \BibitemOpen
  \bibfield  {author} {\bibinfo {author} {\bibfnamefont {Nils}\ \bibnamefont
  {Schöneberg}}, \bibinfo {author} {\bibfnamefont {Julien}\ \bibnamefont
  {Lesgourgues}}, \ and\ \bibinfo {author} {\bibfnamefont {Deanna~C.}\
  \bibnamefont {Hooper}},\ }\bibfield  {title} {\enquote {\bibinfo {title} {The
  {BAO}+{BBN} take on the hubble tension},}\ }\href {\doibase
  10.1088/1475-7516/2019/10/029} {\bibfield  {journal} {\bibinfo  {journal}
  {Journal of Cosmology and Astroparticle Physics}\ }\textbf {\bibinfo {volume}
  {2019}},\ \bibinfo {pages} {029--029} (\bibinfo {year} {2019})}\BibitemShut
  {NoStop}%
\bibitem [{\citenamefont {Scolnic}\ \emph {et~al.}(2018)\citenamefont {Scolnic}
  \emph {et~al.}}]{Pan-STARRS1:2017jku}%
  \BibitemOpen
  \bibfield  {author} {\bibinfo {author} {\bibfnamefont {D.~M.}\ \bibnamefont
  {Scolnic}} \emph {et~al.} (\bibinfo {collaboration} {Pan-STARRS1}),\
  }\bibfield  {title} {\enquote {\bibinfo {title} {{The Complete Light-curve
  Sample of Spectroscopically Confirmed SNe Ia from Pan-STARRS1 and
  Cosmological Constraints from the Combined Pantheon Sample}},}\ }\href
  {\doibase 10.3847/1538-4357/aab9bb} {\bibfield  {journal} {\bibinfo
  {journal} {Astrophys. J.}\ }\textbf {\bibinfo {volume} {859}},\ \bibinfo
  {pages} {101} (\bibinfo {year} {2018})},\ \Eprint
  {http://arxiv.org/abs/1710.00845} {arXiv:1710.00845 [astro-ph.CO]}
  \BibitemShut {NoStop}%
\end{thebibliography}%

\pagebreak

\appendix

\section{Comparison to previous SPT-3G 2018 TEEE}
\label{app:old}

\begin{figure}
    \centering
    \includegraphics[width=0.8\columnwidth]{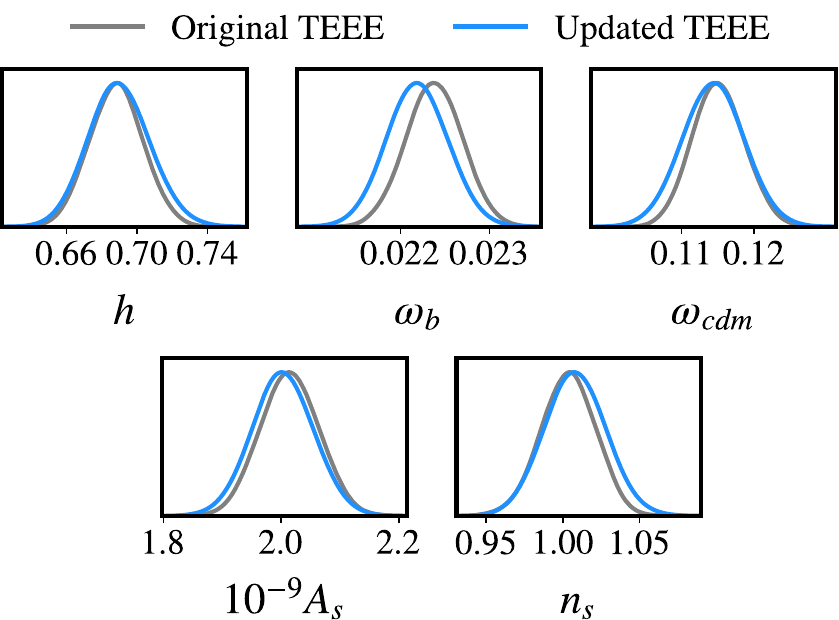}
    \caption{A comparison between the first SPT-3G 2018 polarization data release (`Original TEEE') \cite{SPT-3G:2021eoc} and the recently released polarization data (`Updated TEEE') \cite{SPT-3G:2022hvq} when fit to $\Lambda$CDM.}
    \label{fig:SPT_old_vs_new_LCDM}
\end{figure}

\begin{figure}
    \centering
    \includegraphics[width=\columnwidth]{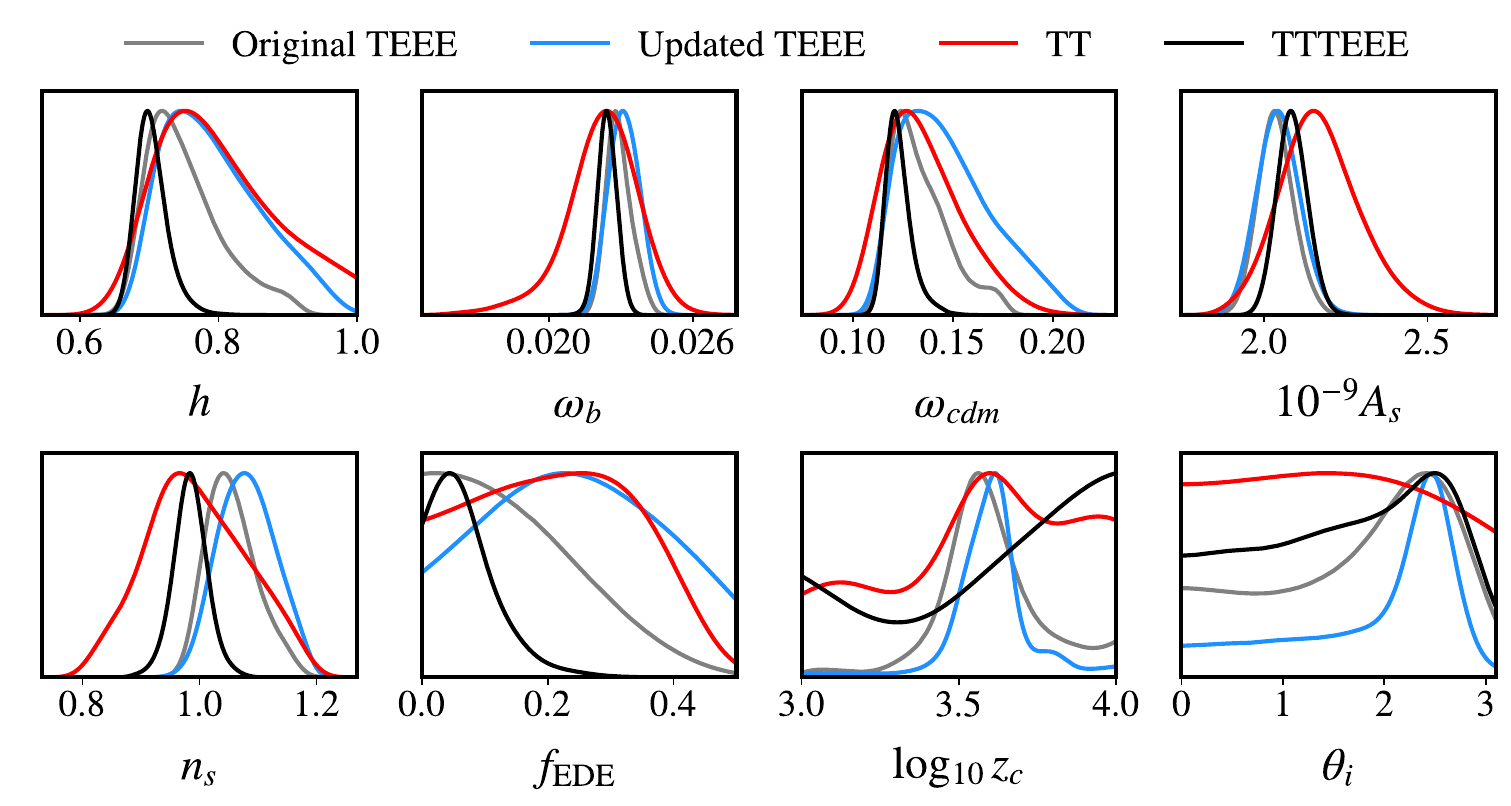}
    \caption{A comparison between the 1D posterior distributions for SPT-3G 2018 temperature and polarization data sets when fit to EDE. We compare the first polarization data release (`Original TEEE') \cite{SPT-3G:2021eoc} the recently released polarization data (`Updated TEEE') \cite{SPT-3G:2022hvq} as well as show the constraints from the temperature power spectrum (`TT') and the full data set (`TT/TE/EE').}
    \label{fig:SPT_old_vs_new}
\end{figure}

In addition to releasing the SPT-3G 2018 temperature power spectrum likelihood, Ref.~\cite{SPT-3G:2022hvq} updated the polarization likelihood. A comparison between the $\Lambda$CDM fit to the original and updated TE/EE SPT-3G 2018 is shown in Fig.~\ref{fig:SPT_old_vs_new_LCDM}. These results are statistically equivalent to those shown in Fig.~13 of Ref.~\cite{SPT-3G:2022hvq} giving us confidence that our MCMC pipeline is working correctly. 

We show a comparison between the 1D posterior distribution for SPT-3G 2018 temperature and polarization data sets when fit to EDE in Fig.~\ref{fig:SPT_old_vs_new}. This figure allows us to compare the first SPT-3G 2018 data release \cite{SPT-3G:2021eoc} to the latest data release \cite{SPT-3G:2022hvq} (i.e., `Original TE/EE' (gray) vs.~`Updated TE/EE' (blue)). Unlike in the case of fitting to $\Lambda$CDM, when fitting to EDE the original and updated TE/EE SPT-3G 2018 likelihoods produce significantly different posterior distributions. Here we can see that the updated TE/EE data set allows for a slightly larger $f_{\rm EDE}$ with a corresponding increase in the allowed values of $h$, $\omega_{cdm}$, and $n_s$. The posterior distribution for $\log_{10} z_c$ is roughly the same, and the posterior for $\theta_i$ is noticably more peaked due, in part, to the shift in $\omega_b$ to slightly larger values. 

\section{Triangle plots and tables}
\label{app:tables}

\begin{figure*}[t]
    \centering
    \includegraphics[width=\textwidth]{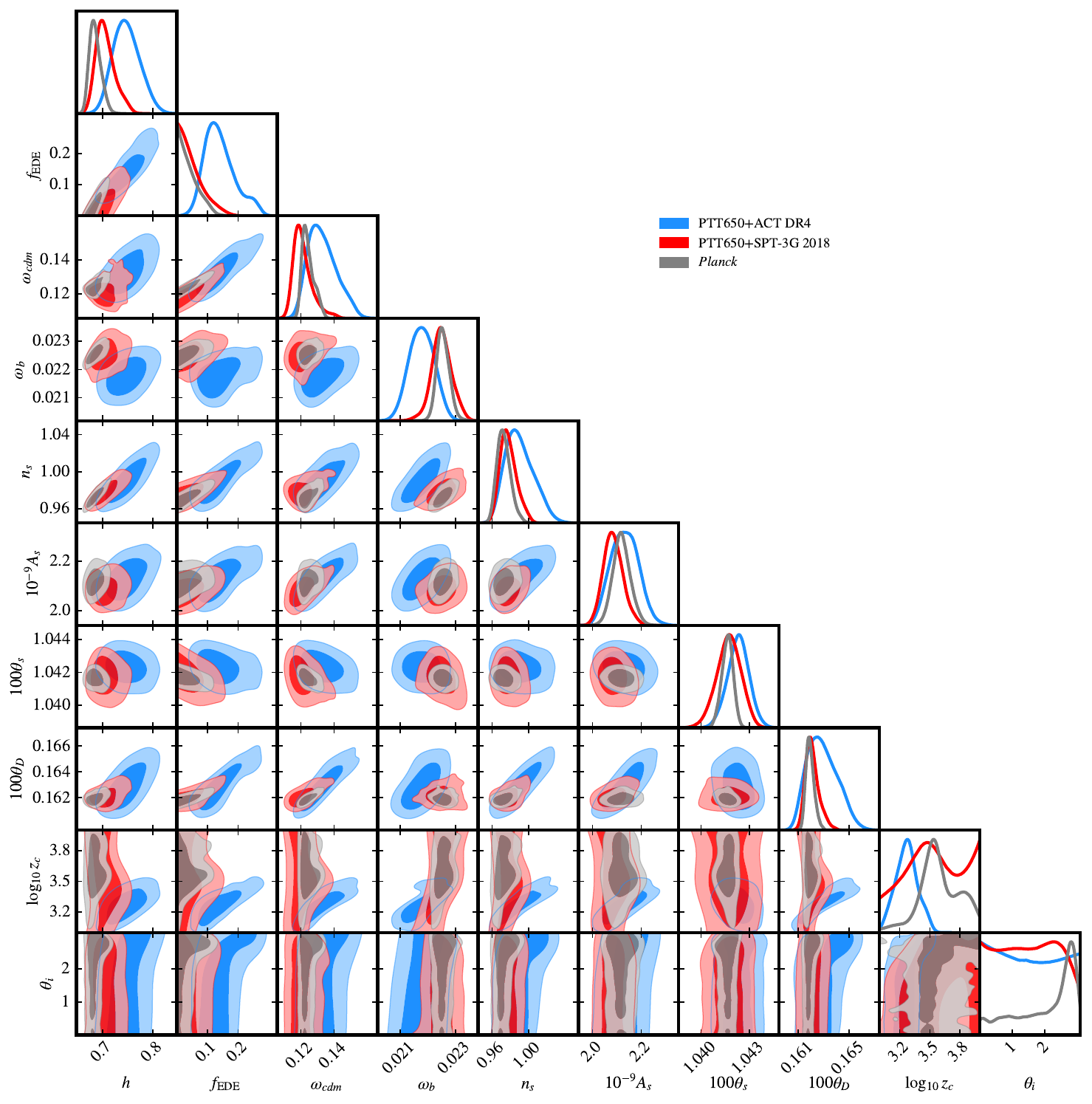}
    \caption{A triangle plot showing the posterior distributions for all of the cosmological parameters when EDE is fit to the combination of PTT650 and ACT DR4 or SPT-3G 2018. }
    \label{fig:big2D}
\end{figure*}

In Fig.~\ref{fig:big2D} we present a triangle plot of all of the cosmological parameters when EDE is fit to the combination of PTT650 and ACT DR4 or SPT-3G 2018. 

In Table \ref{tab:EDE_MCMC_withExt} we give the constraints to the cosmological parameters when fitting a variety of CMB data sets in combination with BAO and Pantheon+.

In Table \ref{tab:EDE_chi2_withExt} we give the best fit $\chi^2$ values for each data set combination shown in Table \ref{tab:EDE_MCMC_withExt}. 

\begin{table*}[h!]
    \centering
        \scalebox{0.75}{
    \begin{tabular}{|c|c|c|c|c|c|c|}
    \hline
    \multicolumn{7}{|c|}{Early Dark Energy}\\
    \hline
    & \multicolumn{2}{|c|}{Planck+Ext} & \multicolumn{2}{|c|}{Planck+SPT-3G 2018+Ext} &  \multicolumn{2}{|c|}{Planck+SPT-3G 2018+ACT DR4+Ext}   \\ 
    \hline
  S$H_0$ES prior?  & no & yes & no & yes & no & yes \\
  
\hline
$h$
	 & $0.6842(0.6942)^{+0.0052}_{-0.011}$ 
	 & $0.7186(0.7212)\pm 0.0078$ 
	 & $0.6823(0.6870)^{+0.0049}_{-0.0088}$ 
	 & $0.7157(0.7192)^{+0.0071}_{-0.0061}$ 
	 & $0.6849(0.6765)^{+0.0051}_{-0.012}$ 
	 & $0.7196(0.7228)\pm 0.0076$ 
	 \\
$f_{\rm EDE}$
	 & $ < 0.083 (0.063) $
	 & $0.128(0.135)^{+0.023}_{-0.021}$ 
	 & $ < 0.071 (0.041)$ 
	 & $0.121(0.131)^{+0.024}_{-0.019}$ 
	 & $< 0.085 (0.009)$ 
	 & $0.131(0.141)\pm 0.021$ 
	 \\
$\log_{10}z_c$
	 &  unconstrained (3.55) 
	 & $3.604(3.568)^{+0.014}_{-0.075}$ 
	 & unconstrained  (3.57)  
	 & $3.569(3.564)^{+0.032}_{-0.041}$ 
	 & unconstrained (3.1)  
	 & $3.543(3.535)\pm 0.030$ 
	 \\
$\theta_i$
	 & unconstrained (2.74)  
	 & $2.73(2.76)^{+0.11}_{-0.090}$ 
	 &  unconstrained (2.70)  
	 & $2.62(2.75)^{+0.18}_{-0.00040}$ 
	 & unconstrained (2.84) 
	 & $2.763(2.791)^{+0.075}_{-0.060}$ 
	 \\
$\omega_{\rm cdm }$
	 & $0.1229(0.1261)^{+0.0013}_{-0.0034}$ 
	 & $0.1329(0.134)\pm 0.0032$ 
	 & $0.1223(0.1238)^{+0.0012}_{-0.0028}$ 
	 & $0.1323(0.1336)\pm 0.0031$ 
	 & $0.1230(0.1203)^{+0.0016}_{-0.0039}$ 
	 & $0.1336(0.1348)\pm 0.0030$ 
	 \\
$10^{2}\omega_{b }$
	 & $2.253(2.256)^{+0.017}_{-0.022}$ 
	 & $2.282( 2.275)\pm 0.022$ 
	 & $2.248(2.249)^{+0.015}_{-0.017}$ 
	 & $2.273(2.270)^{+0.016}_{-0.019}$ 
	 & $2.238(2.225)^{+0.017}_{-0.015}$ 
	 & $2.257(2.253)\pm 0.016$ 
	 \\
$10^{9}A_{s }$
	 & $2.113(2.127)^{+0.029}_{-0.032}$ 
	 & $2.152(2.155)\pm 0.031$ 
	 & $2.105(2.111)\pm 0.030$ 
	 & $2.141(2.145)\pm 0.028$ 
	 & $2.116(2.101)^{+0.028}_{-0.032}$ 
	 & $2.145(2.145)^{+0.029}_{-0.026}$ 
	 \\
$n_{s }$
	 & $0.9713(0.9760)^{+0.0045}_{-0.0074}$ 
	 & $0.9902(0.9915)\pm 0.0059$ 
	 & $0.9705(0.9737)^{+0.0042}_{-0.0063}$ 
	 & $0.9877(0.9895)\pm 0.0051$ 
	 & $0.9724(0.9671)^{+0.0055}_{-0.0086}$ 
	 & $0.9892(0.9901)^{+0.0049}_{-0.0055}$ 
	 \\
$\tau_{\rm reio }$
	 & $0.0549(0.0548)\pm 0.0070$ 
	 & $0.0561(0.0552)\pm 0.0073$ 
	 & $0.0531(0.0533)\pm 0.0070$ 
	 & $0.0545(0.0537)\pm 0.0067$ 
	 & $0.0522(0.0523)\pm 0.0071$ 
	 & $0.0511(0.0502)^{+0.0074}_{-0.0065}$ 
	 \\
  \hline
$S_8$
	 & $0.835(0.841)\pm 0.011$ 
	 & $0.847(0.85)\pm 0.012$ 
	 & $0.833(0.835)\pm 0.011$ 
	 & $0.845(0.847)^{+0.013}_{-0.011}$ 
	 & $0.835(0.828)\pm 0.011$ 
	 & $0.847(0.848)\pm 0.011$ 
	 \\
$\Omega_{m }$
	 & $0.3120(0.3098)\pm 0.0056$ 
	 & $0.3029(0.3027)\pm 0.0049$ 
	 & $0.3124(0.3114)\pm 0.0053$ 
	 & $0.3039(0.3034)\pm 0.0047$ 
	 & $0.3114(0.3128)\pm 0.0052$ 
	 & $0.3028(0.3024)\pm 0.0049$ 
	 \\
\hline 
    $Q_{\rm DMAP}$ & \multicolumn{2}{|c|}{2.6$\sigma$} & \multicolumn{2}{|c|}{2.9$\sigma$} & \multicolumn{2}{|c|}{1.6$\sigma$} \\
    \hline

\end{tabular}
}
\caption{Constraints to the cosmological parameters when various data sets are fit to EDE. In addition to the CMB data set listed in the table all analyses include BAO and Pantheon+ data (denoted by `+ Ext'). The bottom row shows the resulting value for the $Q_{\rm DMAP}$ tension.}
\label{tab:EDE_MCMC_withExt}
\end{table*}

\begin{table*}[h!]
\def\arraystretch{1.2}
\scalebox{1}{
\begin{tabular}{|l|c|c|c|c|c|c|}
    \hline
    \multicolumn{7}{|c|}{EDE} \\
    \hline
    {\emph{Planck}}~high$-\ell$ TT/TE/EE & 2343.3  & 2346.1 & 2344.1& 2346.1& 2350.2 &   2348.9  \\
    {\emph{Planck}}~low$-\ell$ EE & 396.1  & 396.1& 395.9& 396.0& 395.8  & 395.8\\
    {\emph{Planck}}~low$-\ell$ TT & 22.1 & 21.0 & 22.3 &21.2 &  22.9& 21.2\\
    {\emph{Planck}}~lensing & 9.4 &10.1 & 9.2 & 10.1& 8.9& 10.3\\
      SPT-3G 2018 &  $-$ & $-$  & 767.3 & 769.0 & 768.5& 771.0\\
      
    ACT DR4\footnote{In the last column, ACT DR4 data are restricted to $\ell >1800$.}& $-$  & $-$  & $-$ & $-$ & 238.6 &  236.0\\
  
    Pantheon+&1411.5  & 1412.7&  1411.3 & 1412.6 &  1411.1 & 1412.8\\
    BOSS BAO low$-z$ & 1.3 & 1.7& 1.2 &1.7 & 1.1& 1.8\\
    BOSS BAO/$f\sigma_8$ DR12& 6.8 & 7.0&6.9 & 6.9& 6.9& 7.0\\
    S$H_0$ES & $-$ & 2.5 & $-$ & 3.2&$-$  &  2.0\\
    \hline
    total $\chi^2_{\rm min}$& 4190.4 & 4197.3& 4958.2& 4966.7&  5204.0& 5206.7\\
    $Q_{\rm DMAP}$ & \multicolumn{2}{|c|}{2.6$\sigma$} & \multicolumn{2}{|c|}{2.9$\sigma$} & \multicolumn{2}{|c|}{1.6$\sigma$} \\
    \hline
\end{tabular}}
\caption{Best-fit $\chi^2$ per experiment (and total) for EDE when fit to different data combinations. Each column corresponds to a different data set combination.}
\label{tab:EDE_chi2_withExt}
\end{table*}

\end{document}